\documentclass[12pt,onecolumn,draftclsnofoot]{IEEEtran}

\usepackage{bm}
\usepackage[cmex10]{amsmath}
\usepackage{amssymb}
\usepackage{amsfonts}
\usepackage{amsthm}
\usepackage{array}
\usepackage{dsfont}
\usepackage{graphicx}
\usepackage{cite}
\usepackage{bbm}
\usepackage{eurosym}
\usepackage{multirow}
\usepackage{mathtools}
\usepackage{color}
\usepackage{caption}
\usepackage{subcaption}
\usepackage{dblfloatfix}  
\usepackage[hyphens]{url}
\usepackage{hyperref}
\usepackage[hyphenbreaks]{breakurl}

\DeclarePairedDelimiter{\ceil}{\lceil}{\rceil}     
\DeclarePairedDelimiter{\floor}{\lfloor}{\rfloor}

\begin{document}

\title{Smart Meter Data Privacy}

\author{
    \IEEEauthorblockN{Giulio Giaconi\IEEEauthorrefmark{1}, Deniz G\"{u}nd\"{u}z\IEEEauthorrefmark{2}, H. Vincent Poor\IEEEauthorrefmark{3}}\\
    \IEEEauthorblockA{\IEEEauthorrefmark{1}\small{BT Labs, Adastral Park, Martlesham Heath, Ipswich, Suffolk, IP5 3RE, UK}}\\
    \IEEEauthorblockA{\IEEEauthorrefmark{2}Imperial College London, Department of Electrical and Electronic Engineering, London, SW7 2AZ, UK}\\
    \IEEEauthorblockA{\IEEEauthorrefmark{3}Princeton University, Department of Electrical Engineering, Princeton, NJ 08544, USA}
}
    
\maketitle

\begin{abstract}
Smart grids (SGs) promise to deliver dramatic improvements compared to traditional power grids thanks primarily to the large amount of data being exchanged and processed within the grid, which enables the grid to be monitored more accurately and at a much faster pace. The smart meter (SM) is one of the key devices that enable the SG concept by monitoring a household's electricity consumption and reporting it to the utility provider (UP), i.e., the entity that sells energy to customers, or to the distribution system operator (DSO), i.e., the entity that operates and manages the grid, with high accuracy and at a much faster pace compared to traditional meters. However, the very availability of rich and high-frequency household electricity consumption data, which enables a very efficient power grid management, also opens up unprecedented challenges on data security and privacy. To counter these threats, it is necessary to develop techniques that keep SM data private, and, for this reason, SM privacy has become a very active research area. The aim of this chapter is to provide an overview of the most significant privacy-preserving techniques for SM data, highlighting their main benefits and disadvantages.
\end{abstract}

\section{The SG Revolution}\label{smartgridrevolution}

The SG refers to the set of technologies that have been developed to replace an increasingly ageing power infrastructure. Thanks to an extensive use of information and communication technologies and to the introduction of two-way communication links between the UPs and the end customers, the SG allows for improved system reliability, better quality of power delivered and more rapid response to outages and thefts. The SG market is rapidly evolving and is fuelled by a rapid penetration of SG technologies all over the world, as well as by extensive investments, ranging from USD 23.8 billion in 2018 to an estimated USD 61.3 billion by 2023, at a compound annual growth rate of 20.9\% \cite{m&m:2017}.

Key to the SG development is the installation of SMs at the households' premises, which allow near real-time power consumption information to be recorded and sent to the UPs or to the DSOs. SMs are the crucial elements in the SG revolution, as they send electricity consumption measurements at a much higher resolution and with a higher accuracy compared to traditional meters. SMs provide benefits for all parties in the SG. UPs are able to better understand and control the needs of their customers, as well as adjust electricity price dynamically according to short-term generation and consumption variations, being able to communicate this information to the consumers instantly. Additionally, the UPs can generate more accurate bills while reducing the need for back-office rebilling, detect energy theft and outages more rapidly, and implement load-shaping techniques. DSOs are able to reduce operational costs and energy losses, improve grid efficiency, system design and distributed system state estimation, and better allocate resources to the current demand. Consumers themselves take advantage of SMs to monitor their consumption in near real-time, leading to better consumption awareness and energy usage management. Moreover, consumers are able to integrate microgeneration and energy storage devices into the grid, detect failing appliances and waste of energy more quickly, notice expected or unexpected activity, as well as migrate more easily between UPs. The SM market growth demonstrates the value of these technologies, which are expected to reach USD 10.4 billion by 2022, with around 88 million SM installations taking place in 2017 \cite{globaldata:2018}. In order to speed up SM adoption, many countries have introduced legislation that enforces SM installations; for example, most European Union countries need to reach 80\% SM penetration rate by 2020 and 100\% by 2022 \cite{EurDir:2009}. 

The very property that allows SMs to deliver a much improved overall performance in the grid management is, however, also a source of concern for the privacy of SM users. In fact, many appliance load monitoring (ALM) techniques have been developed to gain insights into consumer behavior, inferring consumers' habits or preferences, and the number of household occupants. These privacy concerns are echoed by consumer associations and the media, and even delayed the SM roll-out in the Netherlands in 2009, which proceeded forward only after the customers were given the possibility to opt out from the SM installation \cite{Cuijpers:2012}. Privacy concerns are further exacerbated by recent legislation, such as the General Data Protection Regulation (GDPR) in Europe \cite{GDPR:2016}, which sets a limit on the collection, use and processing of personal information. In particular, article 6 of the GDPR clearly states that the user must give explicit consent to processing of her personal information, and such processing should be limited to only a specific purpose.

\section{ALM Techniques}\label{sec:ALM}

ALM methods are aimed at monitoring a household's power consumption in order to achieve a wide range of benefits for the occupants by providing energy consumption data analysis at the appliance level. ALM techniques provide near real-time feedback on the user's power consumption behavior, are able to detect more power-hungry devices, and allow the automation of demand-side management \cite{Zoha:2012}. Recent improvements in artificial intelligence, data communication and sensing technologies have made the SM benefits even more evident. ALM techniques can be divided into non-intrusive and intrusive load monitoring (NILM and ILM, respectively) techniques. While ILM techniques need to monitor power consumption at multiple points in a household \cite{Ridi:2014}, NILM techniques aim at recognizing the operation of electric loads within a household without the need to physically monitor each electrical device separately, relying only on aggregate SM measurements. ILM techniques are generally more accurate than NILM ones, however, they are also more invasive and expensive to deploy. For this reason, most of the works analyzing privacy for SM users are focused on NILM techniques, which create the biggest concern from a privacy point of view as they can be run by using a single probe attached to the SM and do not need any device to be physically installed within a target household.

The first NILM prototypes were devised in the 80s by George Hart \cite{Hart:1985}. Since then, NILM techniques have evolved in various directions, e.g., by considering either low or high-frequency measurements; by focusing on detecting on/off events by means of either steady state or transient signatures; or by analyzing the raw SM readings, e.g., by studying their frequency spectra. Additionally, both supervised and unsupervised machine learning models have been used for pattern recognition on SM data. Extensive surveys of NILM techniques are provided in \cite{Zoha:2012}, whereas \cite{Ridi:2014} discuss ILM techniques in detail.

\section{SM Privacy Concerns and Privacy-Preserving Techniques}

\begin{figure}[t]
\centering
\includegraphics[width=1\columnwidth]{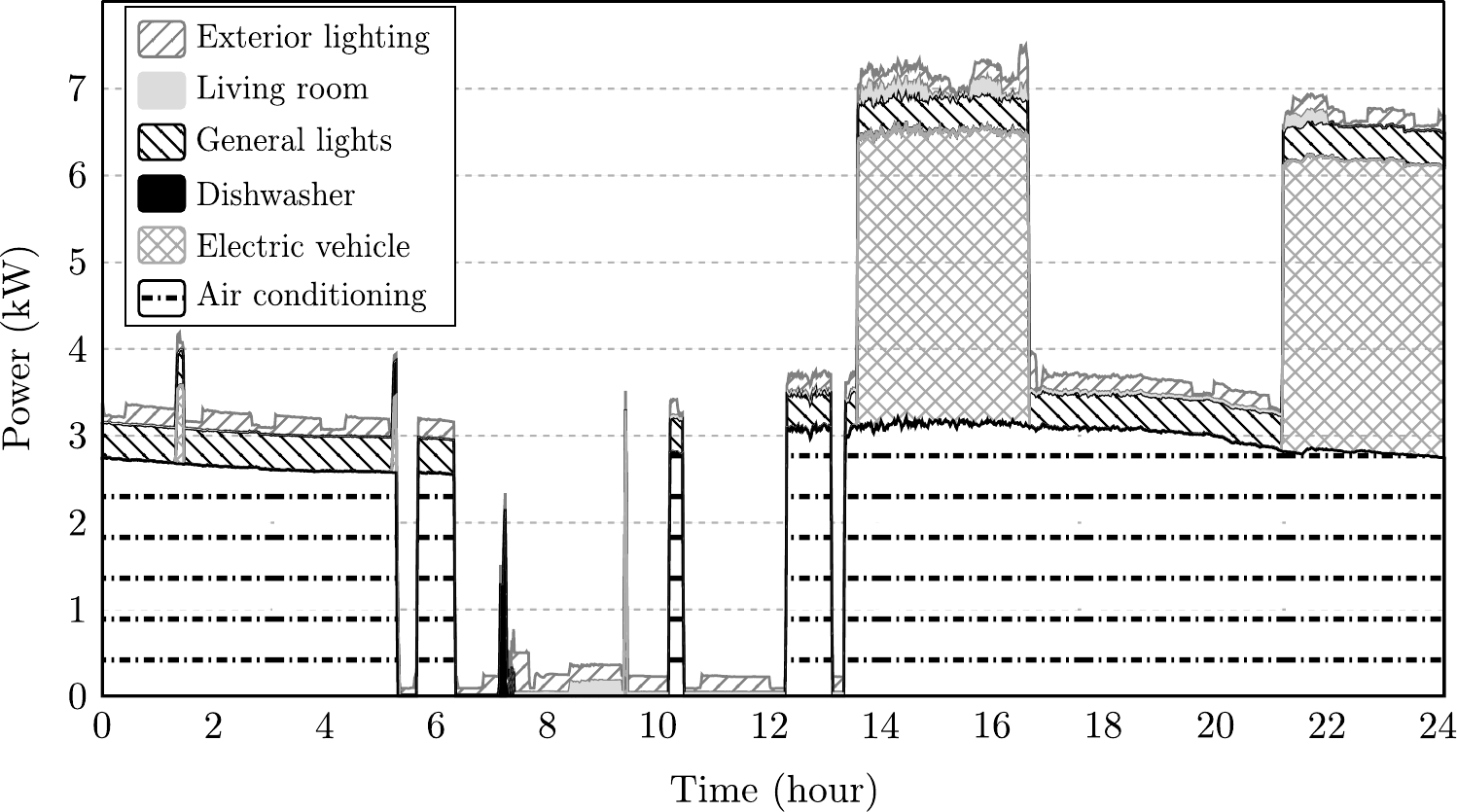}
\caption{A power consumption profile where the consumption of some appliances is highlighted \cite{Giaconi:2018} (data retrieved from the Dataport database \cite{pecanstreet}).}
\label{fig:NILMexample}
\end{figure}

Adoption of SMs and the use of NILM algorithms generate growing concerns about consumer privacy. An example of a typical power consumption profile along with some detected appliances is illustrated in Figure \ref{fig:NILMexample}. It is noteworthy that such precise and accurate information would be available in principle only if very high frequency SM readings were available to an attacker. However, even with low frequency SM readings, the attacker may still be able to gain insights into users' activities and behaviors, determining, for example, a user's presence at home, her religious beliefs, disabilities and illnesses  \cite{Prudenzi:2002, Quinn:2009, Rouf:2012}. Moreover, SM privacy risks could be particularly critical for businesses, e.g., factories and data centers, as their power consumption profiles may reveal sensitive information about the state of their businesses to their competitors. Such important privacy concerns in the use of SMs has raised significant public attention and they have been highly debated in the media and by politicians, and, if not properly addressed, they could represent a major roadblock for this multi-billion dollar industry.

In the following, we adopt the classification introduced in \cite{Giaconi:2018}, and divide the privacy-preserving techniques into \textit{SM data manipulation} (SMDM) techniques, which manipulate SM readings before reporting them to the UP, and \textit{user demand shaping} (UDS) techniques, which modify the actual electricity consumption by shaping it by means of physical devices such as renewable energy sources (RESs) or rechargeable batteries (RBs) \cite{Giaconi:2018}. The main difference between these sets of techniques is that while the SMDM techniques report corrupted or incomplete electrical consumption data to the UP to preserve user's privacy, the UDS techniques report a fully correct measurement, which is, however,  generated by appropriately filtering the original user consumption. Hence, the UDS techniques do not typically suffer from the issue of data mismatch between the UP and the consumer, as opposed to some SMDM techniques, e.g., data obfuscation techniques. SMDM techniques have other notable disadvantages compared to UDS techniques, e.g., an eavesdropper may still be able to measure a user's consumption by installing additional probes outside the target household, hence choosing not to rely exclusively on the SM measurements; or the introduction of trusted third parties (TTPs), considered by many SMDM approaches, which only shifts the problem of trust from the UPs to the TTPs \cite{Giaconi:2018}. Finally, the UDS techniques allow the UP to have full visibility of user's consumption data, as opposed to some SMDM techniques, e.g., data aggregation, anonymization and sharing avoidance techniques. As a result, UDS techniques do not impact the utility of the SG as the actual consumption data is always shared with all the relevant stakeholders. 

On the other hand, the major disadvantage of UDS techniques is that they require the presence of a physical device at the household, which can be costly for the user to purchase and install, such as in the case of RESs and RBs. However, such devices are becoming increasingly available \cite{Munsell:2019}, thanks to government incentives and decreasing cost of solar panels \cite{Goksin:2018}, residential RBs, as well as RBs for electric vehicles \cite{Nykvist:2015}. It is noteworthy that sharing RBs and RESs among multiple users, e.g., within the same neighborhood or block of apartments, results in reduced installation and operation costs as well as allowing management of the available energy in a centralized way, leading to a more efficient use of the available resources among multiple users. Other disadvantages of UDS techniques are that the use of physical sources may impact dynamic pricing and demand response, and such interaction has not been properly investigated yet. Finally, the shaping algorithms of the UDS techniques may prevent detecting anomalous consumption patterns.

Many surveys on SM privacy exist to date, each focusing on different aspects and techniques. Within the SMDM techniques, \cite{Erkin:2013} provides an overview of data aggregation techniques, whereas \cite{Asghar:2017} presents an excellent overview of cryptographic techniques and a wide discussion on privacy requirements and privacy legislation. The earlier survey in \cite{Finster:2015}  discusses mostly SMDM techniques, whereas the recent magazine article \cite{Giaconi:2018} provides a wide review of UDS techniques. Differently from the previous surveys, the focus of this chapter is to provide an up-to-date technological review of the most significant SMDM and UDS techniques, without focussing on legal and normative aspects.

The following analysis considers the possible compromise of SM data, whereas the SM itself is assumed to be tamper-resistant and trusted as it is equipped with a trusted platform module (TPM) to store cryptographic keys and to perform cryptographic operations. However, it is noteworthy that SMs suffer from physical attacks as well, which can be carried out to manipulate consumption data or to steal energy, and which can lead to devastating effects such as explosions \cite{stay_energy_safe}.

The remainder of this chapter is organized as follows. SMDM techniques are analyzed in Section \ref{sec:SMDM}, whereas UDS techniques are discussed in Section \ref{sec:UDS}. Conclusions are drawn in Section \ref{sec:conclusions}.

\section{SMDM Techniques}\label{sec:SMDM}

The main SMDM techniques are data aggregation, obfuscation, anonymization, data sharing prevention and down-sampling. Although these techniques are explained below in distinct paragraphs for simplicity, many of the described works actually consider various combinations of these techniques. In the following, we denote random variables and their realizations by upper case and lower case letters, respectively.  Let $X_{i,t}$ denote the total power requested by the appliances in a household $i$ at time $t$, called the \textit{user load}; and let $Y_{i,t}$, called the \textit{grid load}, denote the electric load that is generated by the application of SMDM techniques to $X_{i,t}$, and which is reported to the UP and the DSO via the SM readings. The objective of the privacy-preserving policies is to keep $X_{i,t}$ private and report only a modified version of it, i.e., $Y_{i,t}$, to the UP. However, in general, the larger the deviation of $Y_{i,t}$ from $X_{i,t}$, the less useful $Y_{i,t}$ is for the UP or the DSO for optimal grid management and correct user billing. For this reason, for these techniques it is often of interest to characterize the trade-off between privacy and utility, e.g., as studied from an information-theoretic point of view in \cite{Sankar:2013TSG}. We remark here that such trade-off is not typically analyzed within the UDS techniques, as the UDS techniques reshape the data by means of physical sources, and report to the UP the power that is actually requested by a household.

\subsection{Data Aggregation Techniques}\label{sec:data_aggregation}

Data aggregation techniques typically propose solutions where SM measurements are encrypted and only the aggregate measurement from $K$ different SMs are revealed to the UP. Aggregation may be achieved with the help of a TTP, which has perfect knowledge of all SM readings and sends only the aggregated measurements to the UP, as proposed in \cite{Bohli:2010}. However, considering a TTP only shifts the problem of trust from one entity (UP) to another (TTP) without actually solving the SM privacy problem itself. Hence, the most significant  data aggregation approaches avoid the presence of a centralized TTP, and propose decentralized approaches where SMs are grouped, e.g., into neighborhoods, and cooperate among themselves to achieve private system operation. Hybrid approaches also exists, where both a TTP and multiple data collectors are considered simultaneously \cite{Petrlic:2010}. 

Aggregation techniques typically require a certification authority that verifies the signatures of single SMs, and the capability of SMs to perform cryptographic operations, e.g., hash functions, symmetric and asymmetric encryption and pseudorandom number generators, which are performed by a TPM \cite{Erkin:2013}. Homomorphic encryption schemes are often used as they allow the UP to perform operations on the cyphertexts of encrypted messages without the necessity of decrypting the messages first, hence keeping the content of the message private. An encryption scheme is said to be homomorphic over an operation $*$ if $\textsf{Enc}(m_1) * \textsf{Enc}(m_2) = \textsf{Enc}(m_1 * m_2)$, $\forall m_1,m_2 \in M$, where $\textsf{Enc}$ denotes the encryption algorithm and $M$ is the set of all possible messages. Homomorphic encryption schemes are either \textit{partial}, e.g., Paillier or ElGamal, which allow only a certain operation to be performed on the cyphertext, or \textit{full}, which allow all operations to be performed but result in high computational complexity \cite{Asghar:2017}. 

Paillier homomorphic encryption and additive secret sharing can be used so that the total power consumption is visible to the UP only at a neighborhood level and every SM in the neighborhood knows only a share of the consumption of all the other SMs \cite{Garcia:2010}. To achieve this, each SM divides its readings into random shares that are encrypted using the public keys of other SMs in the neighborhoods. Then, the UP aggregates and encrypts the readings by means of homomorphic encryption and sends the readings back to the SMs whose public keys were used for encryption. Finally, each SM decrypts the shares encrypted using its public key, adds its own share, and sends the aggregated reading back to the UP. Despite its simplicity, this technique is not scalable as the amount of data increases due to the random shares, and the number of homomorphic encryptions is $\mathcal{O}(K^2)$ \cite{Erkin:2013}. Noise $N_{i,t}$ may be added to individual SM readings to obtain $Y_{i,t} = X_{i,t} + N_{i,t}$, where the noise is computed so that it cancels out once all the readings from all the SMs in a neighborhood are aggregated by the UP, i.e., $\sum_{i=1}^K Y_{i,t} = \sum_{i=1}^K X_{i,t}$ \cite{Kursawe:2011}. Alternatively, each SM may output $g_i^{X_{i,t}+N_{i,t}}$, where $g_i$ is the hash of a unique identifier and $N_{i,t}$ is computed so that they cancel out when the readings are aggregated, as proposed in \cite{Kursawe:2011}, where $g_i$ and $N_i$ are derived by using the Diffie-Hellman key exchange protocol and a bilinear map. However, for the UP to be able to compute the aggregation, it needs to know $g_i$, $\forall i$, and an approximation of the total consumption, and, moreover, this technique results in $\mathcal{O}(K^2)$ messages, $\mathcal{O}(K)$ modulo multiplications and $\mathcal{O}(1)$ exponentiations \cite{Erkin:2013}. As SM data is inherently very rich and multi-dimensional, some techniques can be used to improve the efficiency of homomorphic systems and reduce the computational and communication overhead \cite{Lu:2012}. A further approach is to allow all SMs in a neighborhood to perform intermediate incremental and distributed aggregation, by constructing an aggregation tree rooted at the UP, and using homomorphic encryption to allow end-to-end secure aggregation so that intermediate aggregations are kept private \cite{Li:2011}. 

Aggregation can also be coupled with differential privacy. A function $f$ is defined to be $\epsilon$-differentially private if, for any datasets $D_1$ and $D_2$, where $D_1$ and $D_2$ differ in at most a single element, and for all subsets of possible answers $S \subseteq \textsf{Range}(f)$, the following condition holds: $p(f(D_1) \in S) \leq e^{\epsilon} \cdot p(f(D_2) \in S)$, where $p$ denotes probability. Hence, differentially private functions produce similar outputs for inputs that differ on one element only \cite{Dwork:2006}. A function can be made differentially private by the addition of Laplacian noise $\mathcal{L}(S(f)/\epsilon)$, where $S(f)$ is the global sensitivity of $f$. A Laplace distribution can be generated by summing independent and identically distributed (i.i.d.) gamma distributed random variables $\mathcal{G}(K, \lambda)$, where $\lambda$ is a scale parameter for the Laplace distribution. Hence, $\mathcal{L}(\lambda) = \sum_{i=1}^K [\mathcal{G}_1(K, \lambda) - \mathcal{G}_2(K, \lambda)]$, where $\mathcal{G}_1(K, \lambda)$ and  $\mathcal{G}_2(K, \lambda)$ are drawn independently from the same gamma distribution, i.e., Laplacian noise can be constructed by subtracting gamma distributed random variables. Hence, to achieve a differentially private operation, each SM adds gamma-distributed noise to its readings, encrypt them, and send the encrypted measurement to the UP in the form of $Y_{i,t} = X_{i,t} + \mathcal{G}_1(K, \lambda) - \mathcal{G}_2(K, \lambda)$. Moreover, SM data, corrupted by Laplacian noise and encrypted, can be further aggregated between groups of SMs \cite{Acs:2011}. 

Lately, blockchain technology has also been applied to provide privacy to SM users, especially in the context of data aggregation techniques. The use of blockchain technology, with its decentralized infrastructure, removes the need for a TTP, and the aggregator, or \textit{miner}, is chosen directly from the users within a neighborhood. The miner transmits only the neighborhood aggregate consumption, and each user may create multiple pseudonyms to hide her identity \cite{Guan:2018}. Blockchain has also been considered to provide privacy for users in the context of energy trading \cite{Aitzhan:2018}.

A general issue with data aggregation techniques is the fact that the UP, or the DSO, is prevented from having a clear real-time picture of a single premise's consumption. This can adversely impact the UP in terms of local grid state estimation using SM readings \cite{Majeed:2012}, fault detection at the local level, and ability to implement dynamic pricing to mitigate peak demands \cite{Tonyali:2016}. Moreover, data aggregation techniques typically suffer from the so-called human-factor-aware attack, whereby an attacker may be able to estimate a user's consumption from the aggregate if she knows for example if the user is, or is not, at home \cite{Jia:2014}. Cryptographic techniques, heavily used in data aggregation approaches, typically suffer from high computational complexity, key distribution issues and overhead, and poor scalability, which prevent practical applicability in an SM setting where computational and bandwidth resources are limited. Additionally, cryptographic techniques are vulnerable to statistical attacks and power analysis \cite{power_analysis}.

\subsection{Data Obfuscation Techniques}\label{sec:data_obfuscation}

Data obfuscation revolves around the introduction of noise in the SM readings, i.e., $Y_{i,t} = X_{i,t} + N_{i,t}$, and many works that propose obfuscation techniques also involve aggregating data. In fact, as described in Section \ref{sec:data_aggregation}, if noise is properly engineered across multiple SMs, the aggregation at the UP allows the noise to be removed from the sum of the readings so that the UP is able to retrieve the total power consumption correctly. Alternatively, a simpler solution is to add noise to each SM independently of other SMs, e.g., by adding noise with a null expected value so that the expected value of the readings per each pricing period does not change, i.e., $\mathbb{E[}Y_{i,t}]=\mathbb{E}[X_{i,t} ]$ and $\textsf{Var}[Y_{i,t}] = \textsf{Var}[N_{i,t}]$, as $X_{i,t}$ is not drawn from a random distribution \cite{Bohli:2010}. The UP is able to retrieve an accurate estimate of the aggregate consumption across a group of SMs thanks to the convergence in the central limit theorem. This, however, requires a large number of SMs, which hinders the practical applicability of this technique. More specifically, the number of SMs that are needed is $\left(\frac{w\cdot v \cdot \textsf{Var}[N_{i,t}]}{d}\right)^2$, where $w$ is the confidence interval width of the UP on the aggregate power consumption, $v$ is the maximum peak power used by the consumer for obfuscation, and $d$ is the allowed average deviation in power consumption at the household \cite{Bohli:2010}.

The amount of obfuscation can be determined directly by the UP, and then distributed across multiple SMs with the help of a lead meter. A TTP receives the vector of obfuscated measurements, sums them, and sends them to the UP, which is able to retrieve the correct aggregation value by subtracting the predetermined amount of obfuscation \cite{Kim:2011}. However, the TTP may represent a bottleneck for network traffic and, if compromised, may lead to the disclosure of the original SM readings. To overcome such security and efficiency issues, multiple TTPs can interact to create obfuscation vectors, which are used by each lead meter in its own subnetwork \cite{Tonyali:2016}. The latter approach, which has also been simulated in an IEEE 802.11s wireless mesh network, improves the overall reliability and efficiency but has the obvious disadvantage of requiring multiple TTPs.

Obfuscation techniques deliberately report incorrect readings to the UP, which creates a discrepancy between power production and consumption, and which prevents the UP from quickly reacting to energy outages and thefts and the DSO from properly managing the SG. As an example of the risks involved with obfuscation techniques consider the optimal power flow problem, i.e., characterizing the best operating levels for electric power plants to meet demands while minimizing operating costs. It has been found that noise injection is positively correlated with the generators' power output, and the locational marginal price on each bus of the grid is mostly influenced by the noise applied at links that are in the bus or immediately adjacent to it \cite{Zang:2017}. This example shows how injecting noise may have the consequence of undermining the utility of the SG.

\subsection{Data Anonymization Techniques}\label{sec:data_anonimization}

Data anonymization is about using pseudonyms to hide SMs' identities. Different pseudonyms for the same SM may be used for various functions, e.g., a pseudonym may be allocated for SM data sent at high frequency, necessary for real-time grid monitoring but more privacy sensitive, whereas another pseudonym may be allocated for SM data sent at low frequency, e.g., for billing purposes, and random time intervals are used to reduce correlation between the use of various pseudonyms \cite{Efthymiou:2010SGC}.  The main problem with this approach is how to link the various pseudonyms to the same SM, which can be trivially achieved by using a TTP \cite{Efthymiou:2010SGC}. A disadvantage of these techniques is that recent advances in machine learning and anomaly detection lead to techniques that can successfully de-anonymize SM data \cite{Jawurek:2011}.

\subsection{Data Sharing Prevention Techniques}

These techniques propose methods to process SM data locally at a household, without the need for the readings to be sent to the UP. Hence, the energy bill is computed directly at the household or on any device trusted by the consumer on the basis of publicly accessible tariffs, while only the final bill is revealed to the UP. The issue of SM data privacy does not arise, since user's data never leaves the household, and there is no need for sensitive data to be stored at the UP premises as well. Zero-knowledge proofs \cite{Goldwasser:1985} are employed so that the UP can verify the integrity of the bill, and SM signatures are used to prove the identity of the sender \cite{Molina:2010}. As an example, Pedersen commitments can be used in the form $\textsf{Commit}(x_{i,t}, r_{i,t})$, where $r_{i,t}$ is generated by using known Pedersen generators. These commitments are sent along with the total energy bill over $T$ time slots (TSs) based on the specific time-of-use (ToU) tariff employed, $C = \sum_{t=1}^{T} x_{i,t} c_{t}$, where $c_t$  is the power cost at TS $t$ \cite{Jawurek:2011plug}. Alternatively, non-interactive zero-knowledge techniques can be used along with Camenisch-Lysyanskaya signatures \cite{Camenish:2003}, which can be applied to more complicated non-linear ToU tariffs, i.e., tariffs that change after exceeding certain consumption thresholds \cite{Rial:2011}.

Data sharing prevention techniques may solve the basic problem of metering for billing, however, they cannot be applied in more dynamic scenarios where energy cost changes quickly over short periods of time based on user demands, or when considering demand side management and demand response. Also, these techniques do not solve the privacy problem when SM data needs to be necessarily shared for grid management purposes, e.g., with the DSO.

\subsection{Data Downsampling Techniques}

Alternatively, it is possible to reduce the user load sampling rate, so that the UP receives less frequent SM readings. However, the less frequent the SM readings, the harder it is for the UP (or the DSO) to accomplish their duties. As the SG scenario can be modelled as a closed-loop between the UP and the consumer, whereby the UP reacts to SM readings via demand response, the aim is to minimize the user load sampling rate whilst allowing the closed-loop properties of the system, e.g., safety, stability and reliability, to hold within acceptable limits \cite{Cardenas:2012}.

\section{UDS Techniques}\label{sec:UDS}

\begin{figure}[!t]
\centering
\includegraphics[width=0.9\columnwidth]{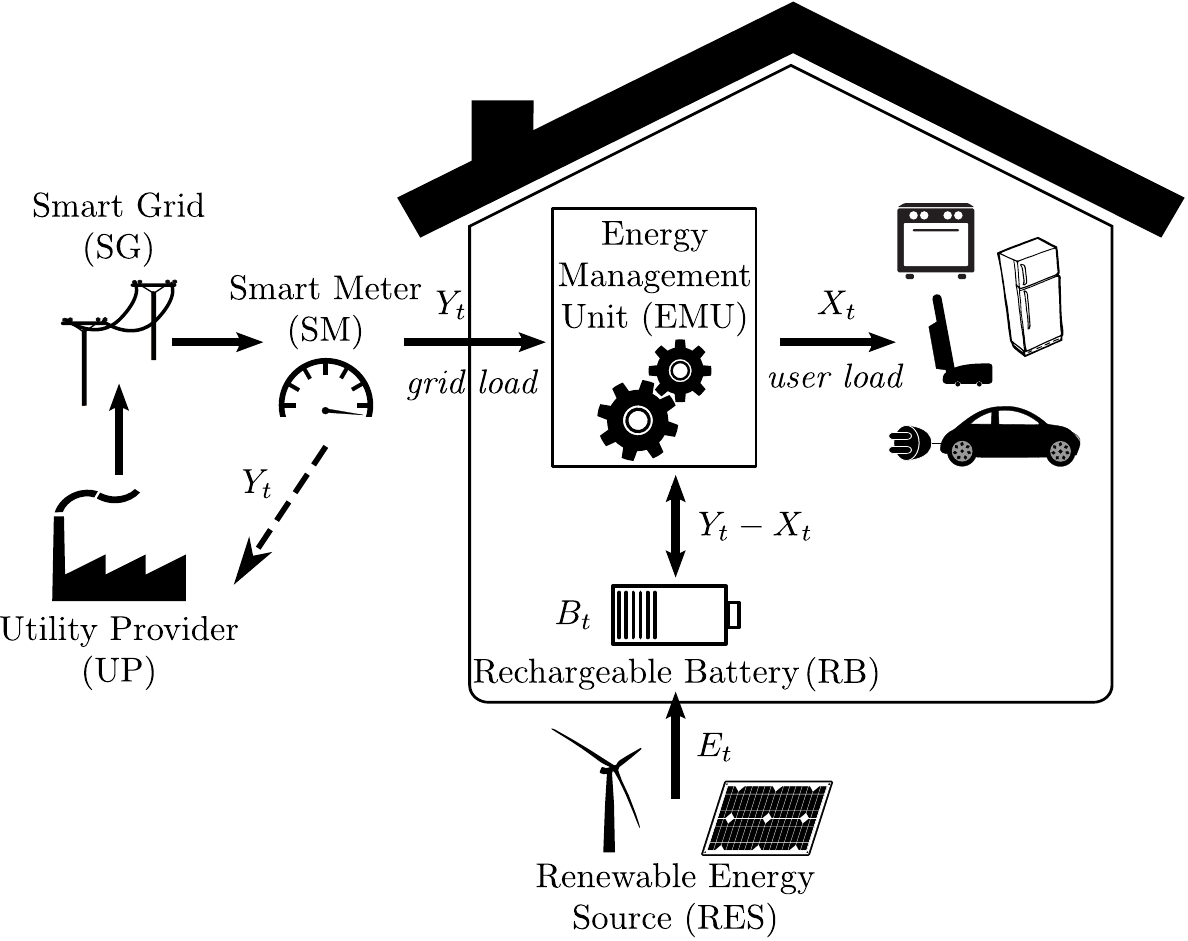}
\caption{Representation of the system model \cite{Giaconi:2018}. $X_t$, $Y_t$, $E_t$ and $B_t$ denote the consumer's energy demand, i.e., the user load, the SM readings, i.e., the grid load, the energy produced by the RES, and the level of energy in the RB at TS $t$, respectively. The meter readings being reported to the UP are shown by the dashed line. The energy management unit (EMU) is the physical or logical unit where the privacy-preserving algorithm resides.}
\label{fig:systemModel}
\end{figure}

Differently from SMDM techniques, UDS techniques report the actual power requested by the consumer, without any manipulation or the addition of any noise. However, what is reported to the UP is not the original load demand of the user, but rather a version of it that is modified by means of the power exchanged with an additional physical device that is present at the household, e.g., an RB or an RES. 

In the following we adopt the same discrete-time SM system model of \cite{Giaconi:2018}, which is represented in Figure \ref{fig:systemModel}. $X_t \in \mathcal{X}$ and $Y_t \in \mathcal{Y}$ denote the total user load and the grid load at TS $t$, respectively, where $\mathcal{X}$ and $\mathcal{Y}$ denote the user load and grid load alphabets, respectively. Each TS duration is normalized to unit time, allowing the use of power and energy values interchangeably within a TS. Also, the user and grid loads are assumed to be constant within a TS, hence representing a discrete-time linear approximation of a continuous load profile, whose accuracy can be arbitrarily increased by reducing the TS duration. Since the aim of the UDS techniques is to protect the privacy of customers from the UP, the DSO, and all the SG parties that may be able to access near real-time power consumption information, the TSs in this model do not correspond to the sampling intervals used for transmitting SM measurements to the UP, but rather to the shorter time intervals that are used to request the actual power from the grid \cite{Giaconi:2018}.

Depending on the user's priorities, part of the demand may not necessarily be satisfied immediately but only by a certain deadline, e.g., fully charging the electric vehicle by 8 a.m., with no specific requirement on the exact time the load needs to take place. Hence, some works explore \textit{load shifting} techniques that allow part of the user load to be shifted to a later time, appropriately called  \textit{elastic demand}, which may be applicable for loads including electric vehicle charging, and dishwasher and clothes washer-dryer cycles. This flexibility allows the consumer to employ \textit{demand response} to increase her privacy as well as to lower the energy cost.

The electricity unit cost at TS $t$, $C_t$, can be modeled as a random variable, or in accordance with a specific ToU tariff, and the total cost incurred by a user to purchase $Y_t$ units of power over a time interval of $\tau_t$ at the price of $C_t$ is thus given by $\tau_t Y_t C_t$. 

\paragraph{Physical Resources: RBs and RESs}

Some of the UDS techniques consider an RB for shaping the grid load, so that the difference between the user and the grid load, $X_t-Y_t$, is retrieved from the RB. The amount of energy stored in the RB at TS $t$ is $B_t \in [0,B_{\max}]$, where $B_{\max}$ denotes the maximum RB capacity, while the RB charging and discharging processes may be constrained by the so-called charging and discharging power constraints $\hat{P}_c$ and $\hat{P}_d$, respectively, i.e., $-\hat{P}_c \leq X_t - Y_t \leq \hat{P}_d$, $\forall t$, and additional losses in the battery charging and discharging processes may be taken into account to model a more realistic energy management system. The battery wear and tear due to charging and discharging can also be considered and modeled as a cost variable \cite{Yang:2015TSG}. Some works also consider a less stringent constraint on the average power that can be retrieved from an RB $\bar{P}$, i.e., $\mathbb{E}\big[\frac{1}{n}\sum_{t=1}^n (X_t - Y_t)\big] \leq \bar{P}$. Where an RESs is considered, the renewable energy generated at TS $t$ is denoted by $E_t\in \mathcal{E}$, where $\mathcal{E}=[0,E_{\max}]$ depending on the type of energy source. The amount of energy in the RB at TS $t+1$, $B_{t+1}$, can be computed on the basis of $B_t$ as
\begin{equation}\label{eq:battery_update}
B_{t+1} = \min \Big\{B_{t} + E_t - (X_t - Y_t), B_{\max} \Big\}.
\end{equation}

Works that characterize theoretical expressions or bounds for the level of privacy achieved in SM systems typically consider the random processes $X$ and $E$ to be Markov or made up of sequences of i.i.d. random variables. Some works also study the scenario where the UP knows the realizations of the renewable energy process $E$, which may occur if, for example, the UP has access to additional information from sensors deployed near the household that measure various parameters, e.g., solar or wind power intensity, and if it knows the specifications of the user's renewable energy generator, e.g., model and size of the solar panel. It is noteworthy that RBs and RESs can be used for both privacy protection and cost minimization, and using them jointly greatly increase the potential benefits. For example, from a cost-saving perspective, the user may be able to use the generated renewable energy when electricity is more expensive to buy from the grid, and may even be able to sell surplus energy to grid.
 
\paragraph{The Energy Management Policy (EMP)}
The EMP $f$, implemented by the EMU, decides on the grid load at any TS $t$ based on the previous values of the user load $X^t$, renewable energy $E^{t}$, level of energy in the battery $B^t$, and grid load $Y^{t-1}$, i.e., 
\begin{equation} \label{eq:startingPolicy}
f_t: \mathcal{X}^t \times \mathcal{E}^t \times \mathcal{B}^t \times \mathcal{Y}^{t-1} \rightarrow \mathcal{Y}, \qquad \forall t,
\end{equation}
where $f \in \mathcal{F}$, and $\mathcal{F}$ denotes the set of feasible policies, i.e., policies that produce grid load values satisfying the RB and RES constraints at any time, as well as the battery update equation (\ref{eq:battery_update}). The EMP is chosen so that it optimizes the user privacy along with other targets, e.g., the cost of energy or the amount of wasted energy, and it has to satisfy the user demand. The EMP in (\ref{eq:startingPolicy}) can be analyzed either as an \textit{online EMP}, which only considers information available causally right up to the current time to make a decision, or as an \textit{offline EMP}, in which case the future user load values are assumed to be known in a non-causal fashion. Although online algorithms are more realistic and relevant for real-world applications, offline algorithms may still lead to interesting intuition or bounds on the performance, and non-causal knowledge of the electricity price process as well of power consumption for large appliances such as refrigerators, boilers, heating and electric vehicles may still be considered valid. 
 
A number of privacy measures and techniques have been proposed in the literature, each with its own advantages and drawbacks. In the following we review the most significant approaches, and distinguish between \textit{heuristic} and \textit{theoretically-grounded} techniques \cite{Giaconi:2018}. Since NILM algorithms look for sudden changes in the grid load profile $y_t - y_{t-1}$, and assign them to specific electric appliances' on/off events, the so-called \textit{features} \cite{Zoha:2012Sensors}, heuristic techniques are aimed at minimizing such changes in the grid load. However, since these approaches counter specific NILM techniques, the validity of their privacy guarantees are also limited only against these attacks, and they do not provide theoretical assurances on the amount of privacy that can be achieved. On the contrary, theoretically-grounded techniques typically provide a rigorous definition of privacy measure, and characterize ways to achieve privacy providing theoretical guarantees under that measure. However, their practical implementation may be harder to achieve and demonstrate.

\subsection{Heuristic Privacy Measures: Variations in the Grid Load Profile}
 
Generating a completely flat (equivalently, deterministic) or a completely random (independent of the user load) grid load profile can provide privacy against NILM algorithms. However, this could be achievable in practice only by having a very large RB or a very powerful RES, or by requesting more power than needed from the UP, both options being potentially extremely costly for the consumer. In the following we describe various EMPs on the basis of the privacy measure or the specific technique being adopted.

\subsubsection{Optimization Techniques}
 
 \begin{figure}[!t]
\centering
\includegraphics[width=0.8\columnwidth]{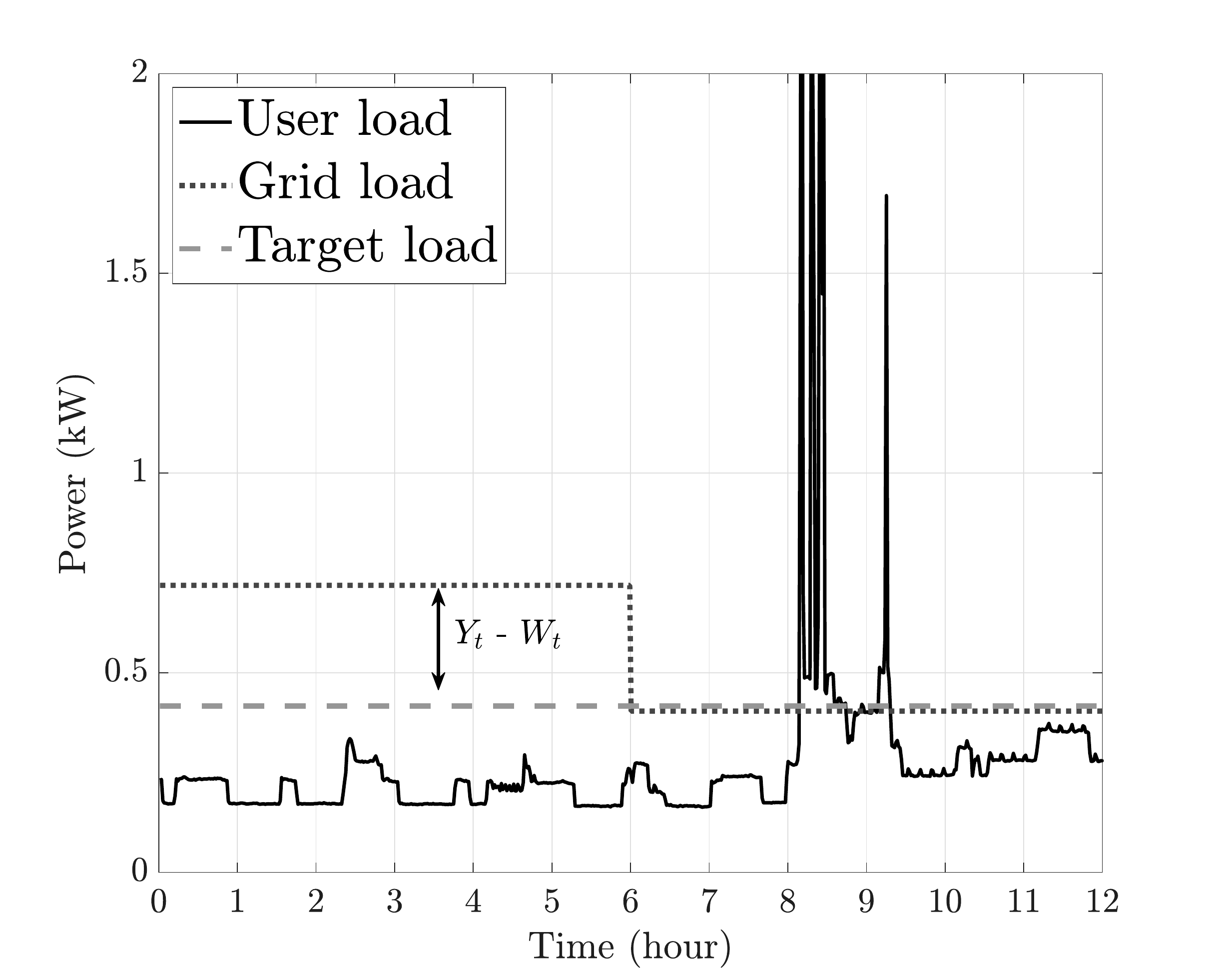}
\caption{Examples of user load, grid load, and target load profiles when considering a constant target load profile \cite{Giaconi:2018}. In this figure the ``distance'' between the grid load and the target load, $Y_t-W_t$, is highlighted. The algorithms presented in this section are aimed at minimizing the average squared distance.}
\label{fig:variance}
\end{figure}

A possible solution to reducing the variations in the grid load profile is to set up an appropriate constant \textit{target load profile} $W$ and try to match it over time. The capability of a privacy-preserving algorithm is then measured by how tight this match is, i.e., how small the variance of the grid load $Y$ is around $W$  \cite{Tan:2017TIFS}: 
\begin{equation}\label{eq:loadVarianceExp}
\mathcal{V}_T \triangleq \frac{1}{T} \sum_{t=1}^{T} \mathbb{E}\Big[ (Y_t-W)^2\Big],
\end{equation}
where the expectation is over $X_t$ and $Y_t$, and $W = \mathbb{E}[X]$ may be considered. In fact, in the limiting scenario where the target load profile is completely flat this would be equivalent to leaking only the average power consumption to the UP, unless more power than that needed by the consumer has been requested. This scenario is shown in Figure \ref{fig:variance}, where the solid line represents the user load, the dashed line represents the constant target load profile, and the dotted line represents the actual grid load profile. Additionally, also the cost of energy, expressed by the following equation, may need to be minimized:
\begin{equation}\label{eq:costExp}
\mathcal{C}_T \triangleq \frac{1}{T} \sum_{t=1}^{T} \mathbb{E}\Big[C_t Y_t\Big].
\end{equation}

A solution to the joint optimization of Eqs. (\ref{eq:loadVarianceExp}) and (\ref{eq:costExp}) can be characterized for an offline framework, where the optimal privacy and cost of energy can be found as the points on the Pareto boundary of the convex region formed by all the cost and privacy leakage pairs by solving the following convex optimization problem \cite{Tan:2017TIFS}:
\begin{equation}\label{eq:Tan_2017_JIFS_target}
\min_{Y_t \geq 0} \sum_{t=1}^T \bigg[(1-\alpha) Y_t C_t + \alpha  (Y_t-W)^2\bigg],
\end{equation}
where $0\leq\alpha\leq1$ strikes the trade-off between privacy and cost of energy, which can be set up by the user. The solution to Eq. (\ref{eq:Tan_2017_JIFS_target}) has a water-filling interpretation with a variable \textit{water level} due to to the instantaneous power constraints. When modelling the battery wear and tear caused by charging and discharging the RB, the optimization can be expressed as \cite{Yang:2015TSG}:
\begin{equation}\label{eq:P1}
\min \frac{1}{T} \sum_{t=1}^T \mathbb{E}\Big[C_t Y_t +\mathds{1}_B(t) C_B + \alpha (Y_t - W)^2 \Big],
\end{equation}
where $\mathds{1}_B(t) = 1$ if the battery is charging/discharging at time $t$, and $0$ otherwise and $C_B $ is the battery operating cost due to the battery wear and tear caused by charging and discharging the RB; and the expectation in (\ref{eq:P1}) is over the probability distributions of all the involved random variables, i.e., $X_t$, $Y_t$, and $C_t$. The solution to Eq. (\ref{eq:P1}) has been characterized for an online setting by means of a Lyapunov function with a perturbed weight and by adopting the \textit{drift-plus-penalty} framework, which consists of the simultaneous minimization of a so-called \textit{drift}, i.e., the difference in the level of energy in the RB at successive time instants, and of a \textit{penalty} function, i.e., the optimization target itself. The solution to this problem leads to a mixed-integer nonlinear program, which can be solved by decomposing the problem into multiple cases and solving each of them separately \cite{Yang:2015TSG}. With a similar approach, it is possible to constrain the grid load to be within a certain maximum range $\lambda$ of an average historical load $\bar{Y}$ at any TS, i.e., $\lambda \leq Y(t) - \bar{Y} \leq \lambda$ \cite{Wu:2016}. In the latter work, load shifting is analyzed to exploit the possibility of shifting non-urgent appliances to improve the privacy-cost trade-off, and an anomaly detection method is developed to detect attacks on the electricity prices publicized to consumers.

\begin{figure}[!t]
\centering
\includegraphics[width=0.8\columnwidth]{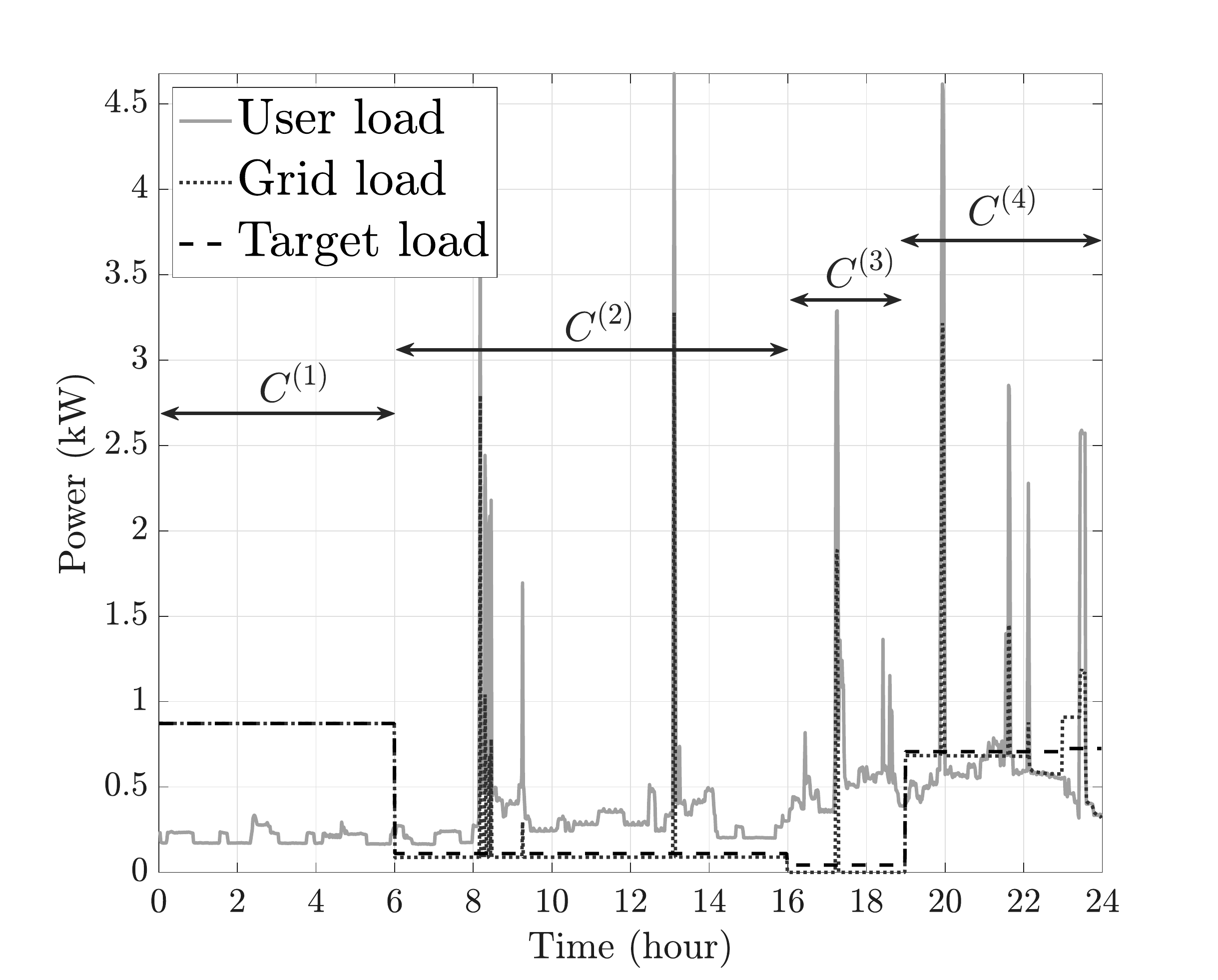}
\caption{Examples of user load, grid load, and target load profiles over the course of a day when considering a piecewise target load profile \cite{Giaconi:2018, Giaconi:2017SGC}. The arrows highlight the various price periods. Note that the target assumes a different constant value for each price period. Electricity consumption data retrieved from the UK-Dale dataset \cite{UKDALE}.}
\label{fig:variance_piece}
\end{figure}

\begin{figure}[!t]
\centering
\includegraphics[width=.8\columnwidth]{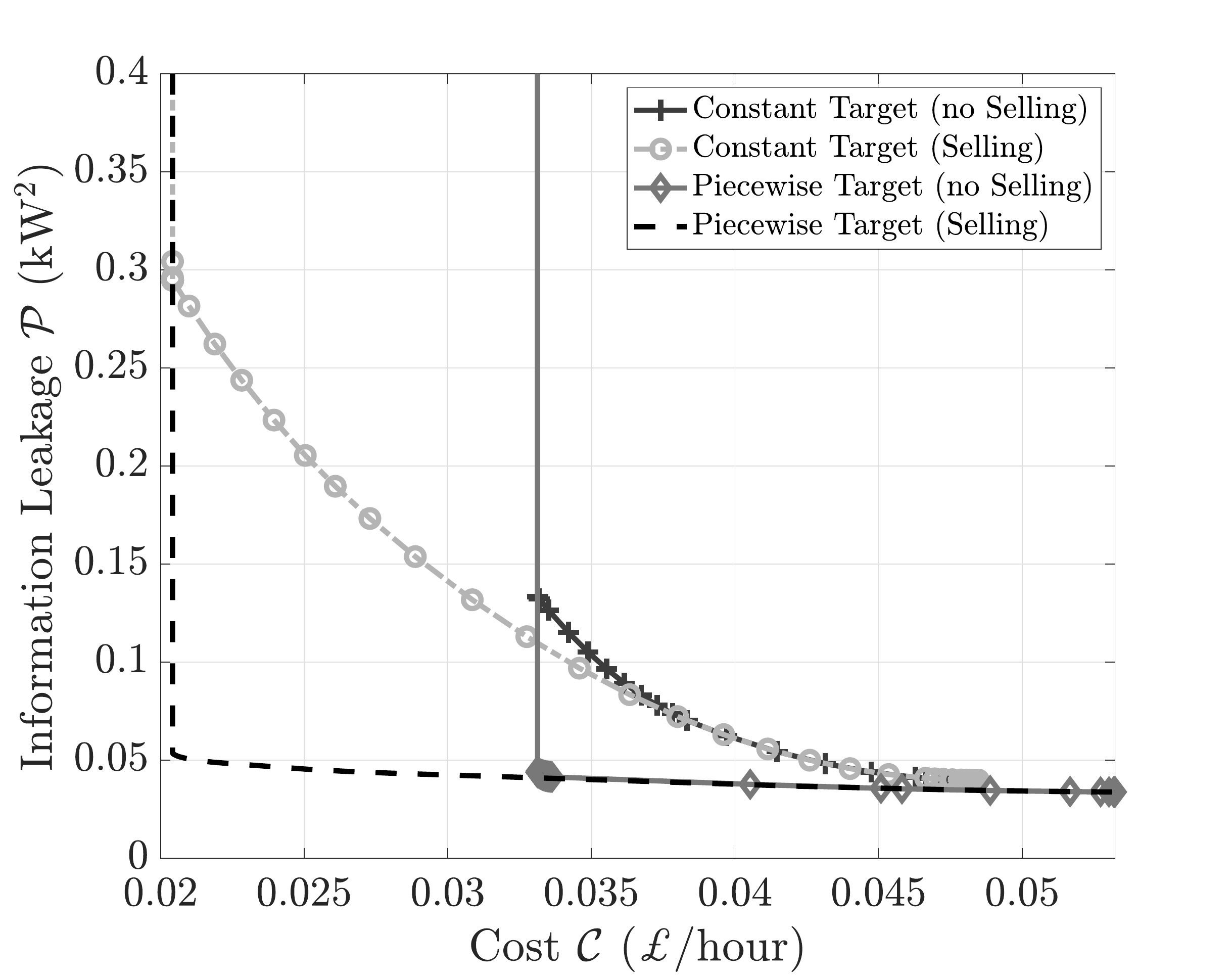}
\caption{Privacy-cost trade-off when using a Powervault G200-LI-4KWH RB \cite{powervault} and adopting the strategies characterized in \cite{Tan:2017TIFS} and \cite{Giaconi:2017SGC}.}
\label{fig:tradeoff}
\end{figure}

Matching a completely constant target load profile is not feasible most of the time as that would require the user to have a large RB or RES. Additionally, it would conflict with the cost saving objective as the constant target load completely disregards any ToU tariff. Instead, it would be reasonable to assume that a user would prefer to request more electricity over less expensive TSs compared to more expensive TSs. To allow such flexibility, one can set a piece-wise constant target load profile, as shown in Figure \ref{fig:variance_piece} \cite{Giaconi:2017SGC}. Accordingly, the optimization problem can be expressed as
\begin{equation}\label{eq:convexOptConstantPiecewise}
\min_{Y_t,W^{(i)}}  \sum_{i=1}^{M} \sum_{t=t_{c^{(i-1)}}}^{t_{c^{(i)}}-1} \Big[  \alpha (Y_t - W^{(i)})^2 + (1-\alpha) Y_t C^{(i)} \Big],
\end{equation}
where $C^{(i)}$ and $W^{(i)}$ are the cost of the energy purchased from the UP and the target profile level during the $i$-th price period, respectively, where $1 \leq i \leq M$, $M$ is the total number of price periods during time $T$, and the $i$-th price period spans from time slot $t_{c^{(i-1)}}$ to $t_{c^{(i)}}$. As expected, considering a piece-wise constant target profile allows the system to reach a better privacy-cost trade-off compared to a constant target profile, as shown in Figure \ref{fig:tradeoff}, and allowing energy to be sold to the grid improves the trade-off even further \cite{Giaconi:2017SGC}. However, it is noteworthy that adopting a piece-wise constant target profile introduces an inherent information leakage compared to a constant target load profile that is not fully captured by the trade-off in Figure \ref{fig:tradeoff}. 

\begin{figure}[!t]
\begin{subfigure}[t]{.49\columnwidth}
\centering
\includegraphics[width=1\columnwidth]{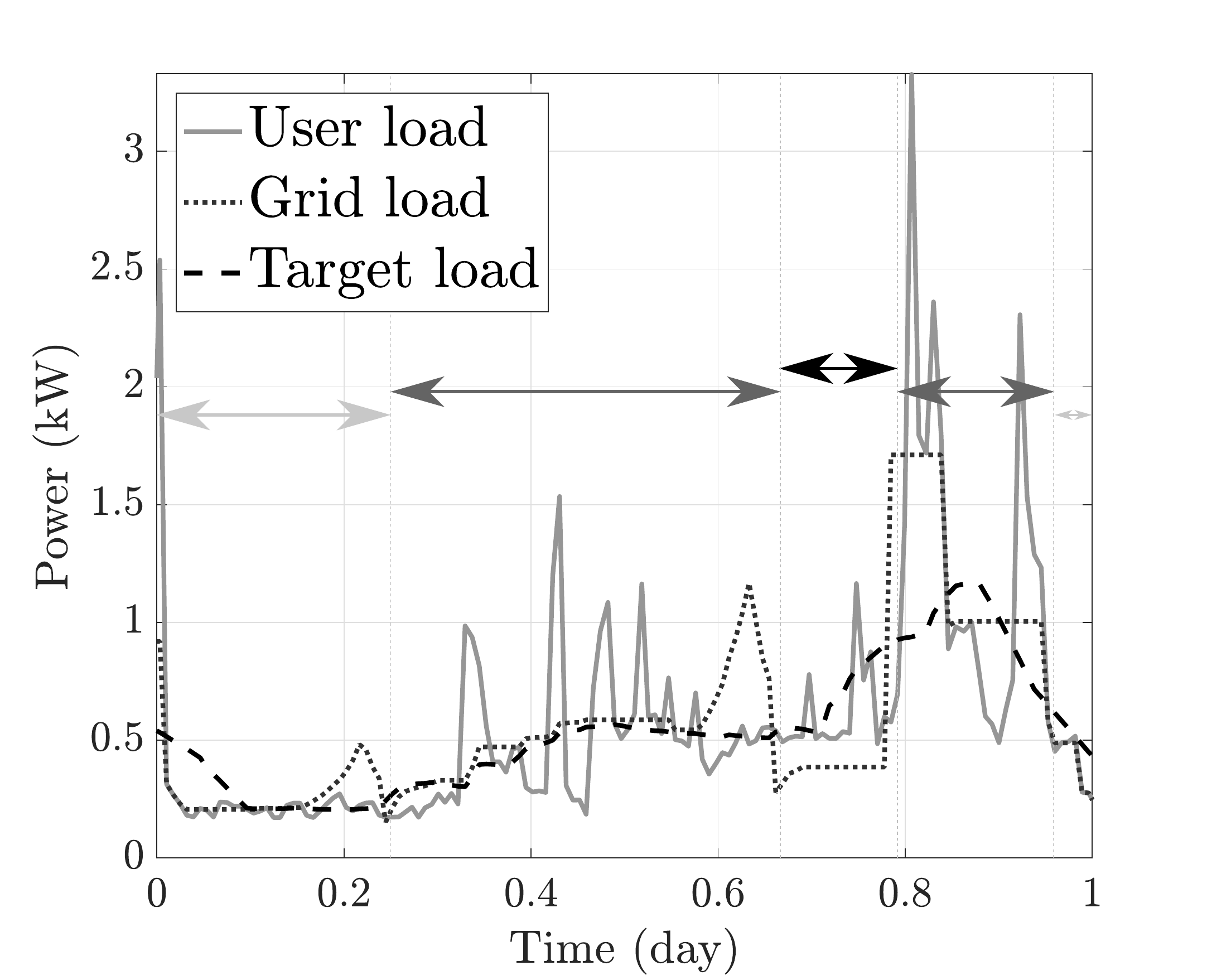}
\caption{SHM, no energy selling.}
\label{fig:constantShort}
\end{subfigure}
\begin{subfigure}[t]{.49\columnwidth}
\centering
\includegraphics[width=1\columnwidth]{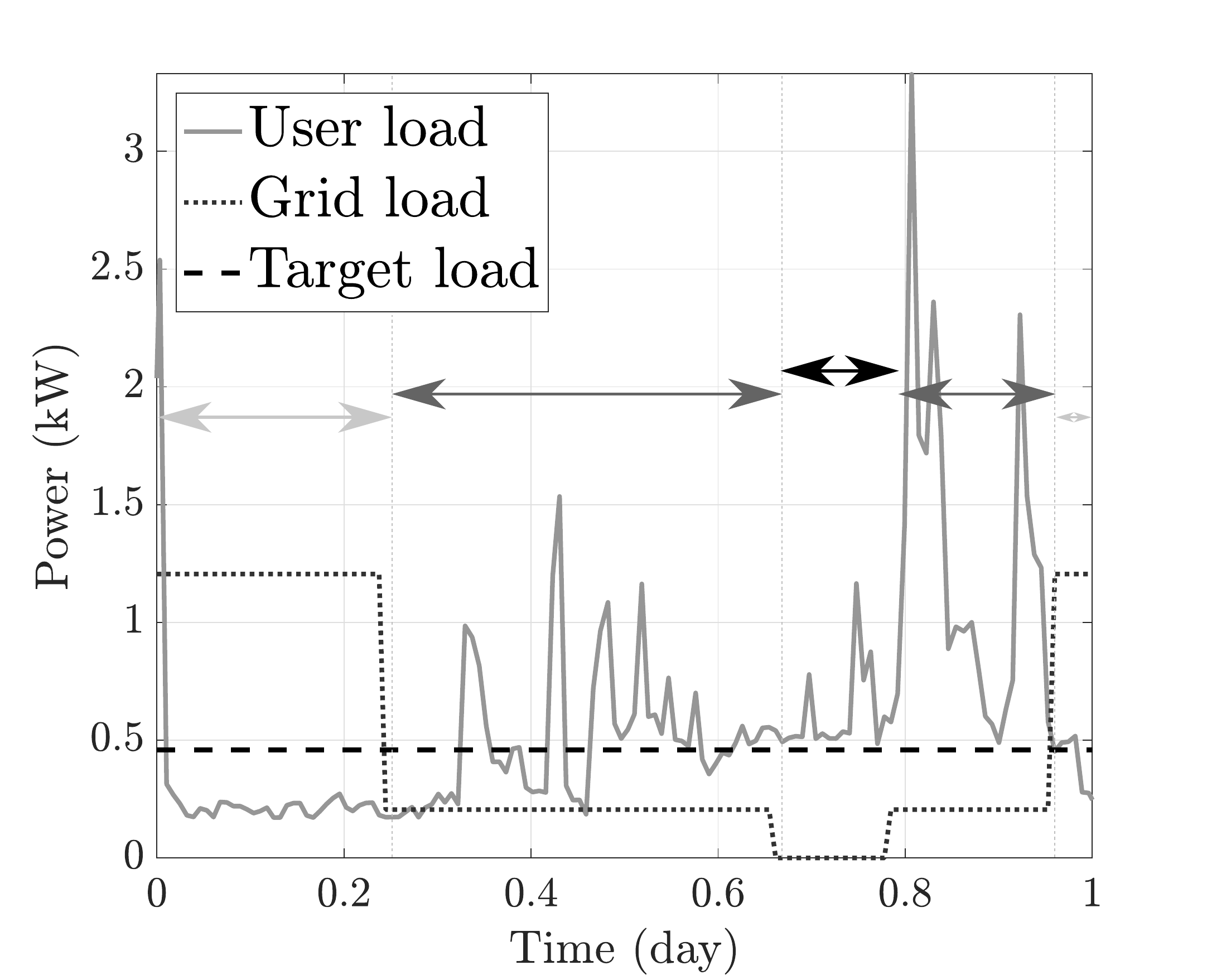}
\caption{LHM, no energy selling.}
\label{fig:constantLong}
\end{subfigure} \\
\begin{subfigure}[t]{.49\columnwidth}
\centering
\includegraphics[width=1\columnwidth]{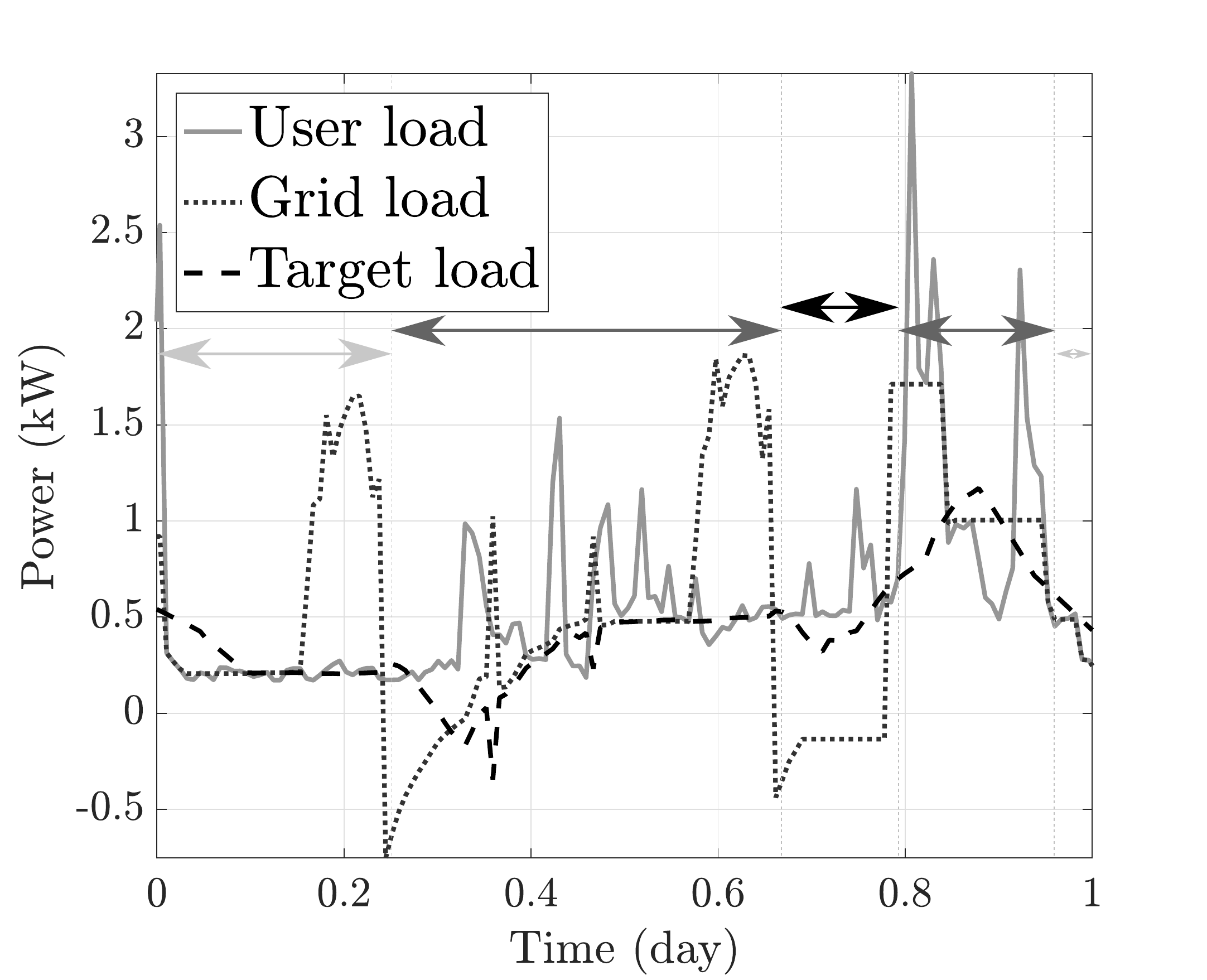}
\caption{SHM, energy selling.}
\label{fig:constantShortS}
\end{subfigure}
\begin{subfigure}[t]{.49\columnwidth}
\centering
\includegraphics[width=1\columnwidth]{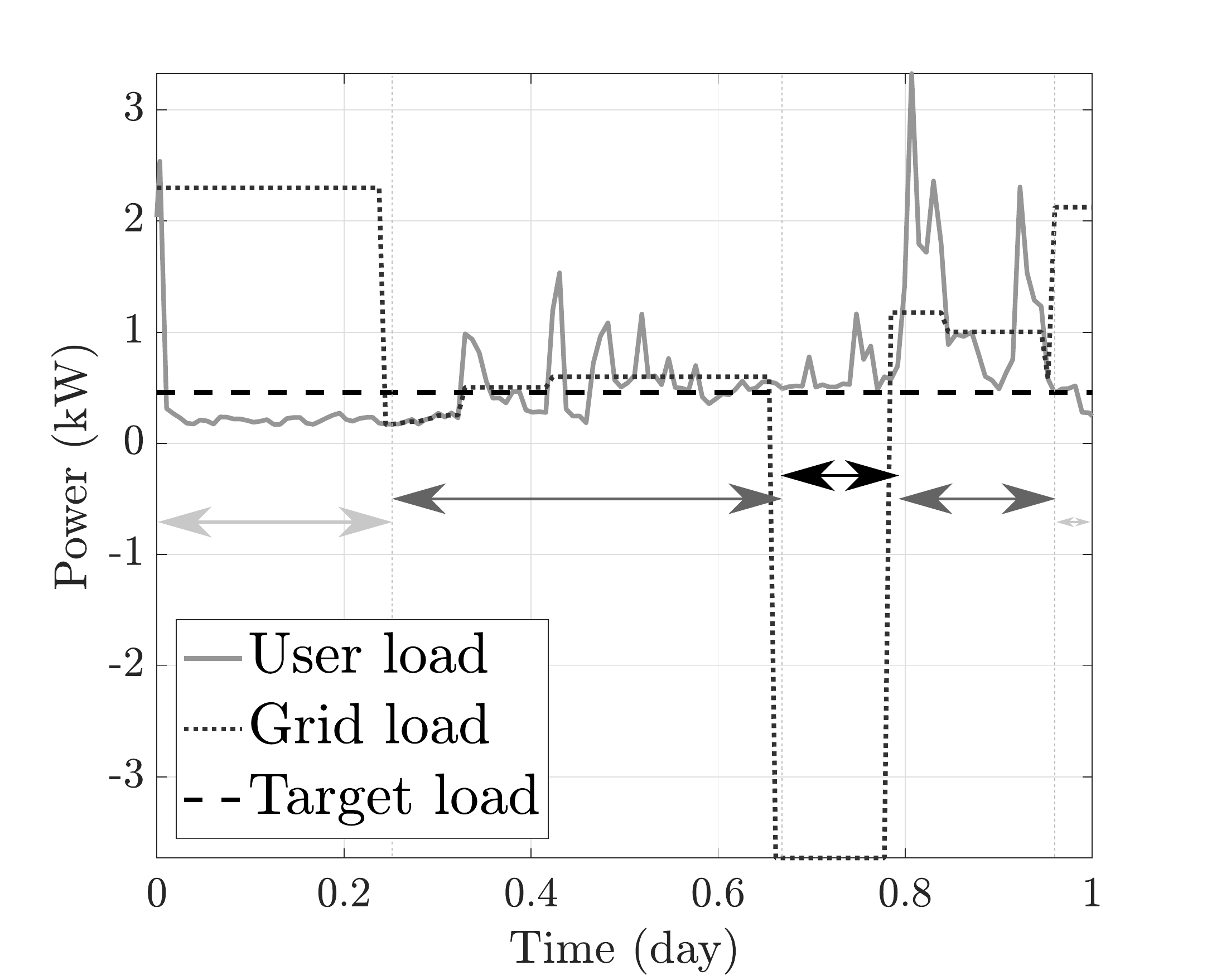}
\caption{LHM, energy selling.}
\label{fig:constantLongS}
\end{subfigure}
\caption{Power profiles for the SHM and the LHM scenarios, $\alpha=0.5$ and $H_F=H_P=2$ hours \cite{Giaconi:2018_draft}.  Off-peak, medium and peak price periods for the electricity cost are denoted by increasingly darker arrows.}\label{fig:constantLongVSShort}
\end{figure}

The adoption of a piece-wise constant target load profile has also been studied in a more realistic scenario, called the \textit{short horizon model} (SHM), in which the consumer's future consumption profile is known to the EMU only for the next $H_F$ TSs, and where a \textit{moving} or \textit{receding horizon} model is considered \cite{Giaconi:2018_draft}. Let $\overline{t+H_F} \triangleq \min\{t+H_F,T\}$, and let $\overline{t-H_P} \triangleq \max\{t-H_P,0\}$. Then, the optimization problem is formulated as
\begin{equation}\label{eq:ConstantShort}
\min_{Y_t^{\overline{t+H_F}},W_t}  \alpha \sum_{\tau=\overline{t-H_P}}^{\overline{t+H_F}} (Y_{\tau} - W_t)^2 + (1-\alpha)\sum_{\tau=t}^{\overline{t+H_F}} Y_{\tau} C_{\tau},
\end{equation}
which states that at TS $t$ the EMP produces the optimal grid load for the current TS and the prediction horizon $Y_t^{\overline{t+H_F}}$, and the optimal target load for the current time $W_t$. It is noteworthy that the SM remembers the consumption that occurred during the previous $H_P$ TSs, considered in the term $\sum_{\tau=\overline{t-H_P}}^{t-1} (Y_{\tau} - W_{t})^2$, to ensure a smooth variation of the overall target load profile. Figure \ref{fig:constantLongVSShort} compares the load profiles of the SHM (Figures \ref{fig:constantShort} and \ref{fig:constantShortS}) and the offline scenario, called the \textit{long horizon model} (LHM) (Figures \ref{fig:constantLong} and \ref{fig:constantLongS}) over the course of one day, also including the scenario where energy can be sold to the grid. The LHM results in a flatter grid load profile compared to the SHM, however, the SHM is also able to flatten the consumption peaks to some extent and the resulting peaks in the grid load are not aligned with the original peaks in the user load \cite{Giaconi:2018_draft}. 

\begin{figure}[!t]
\begin{subfigure}[t]{.5\columnwidth}
\centering
\includegraphics[width=1\columnwidth]{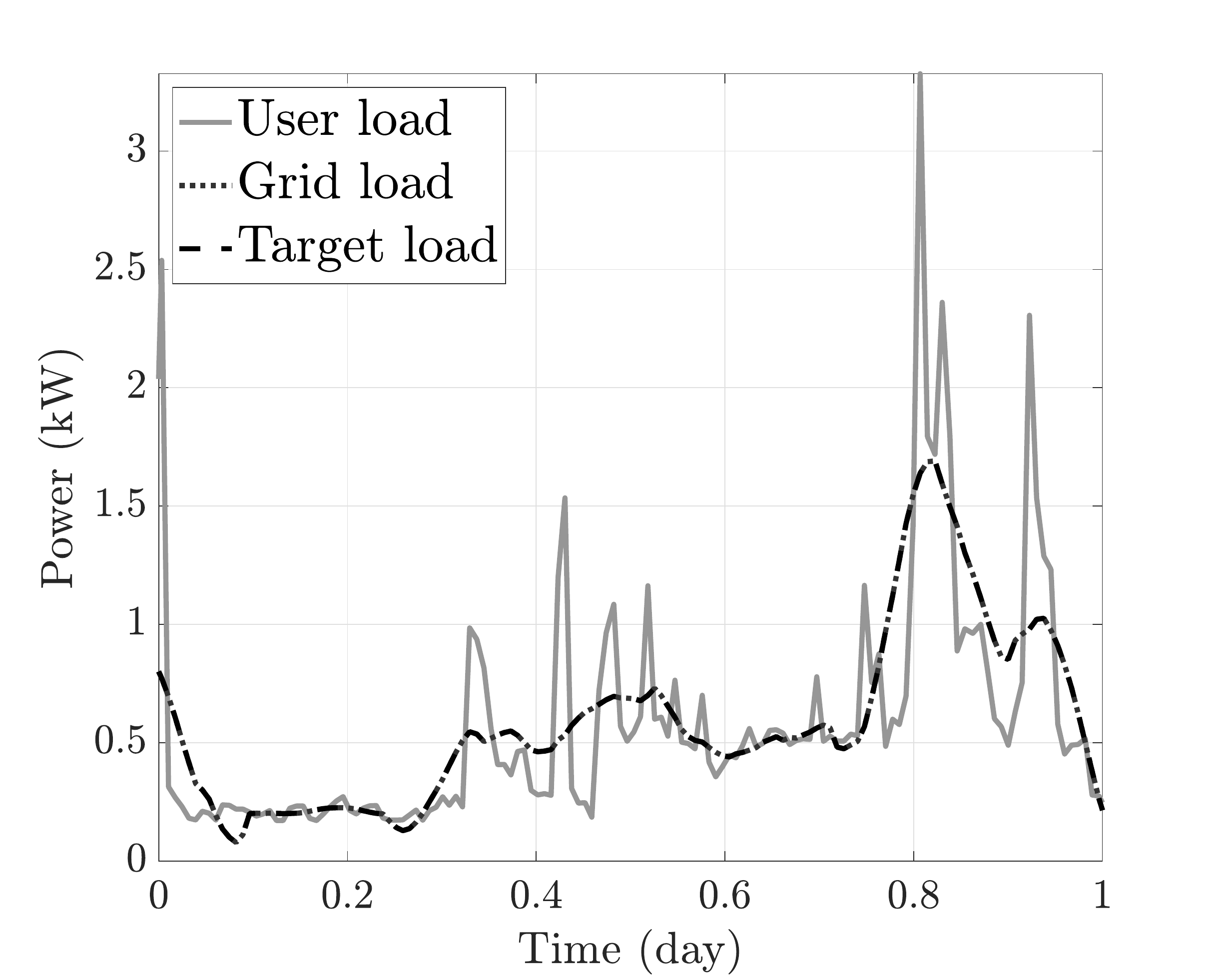}
\caption{SHM.}
\label{fig:filterShort}
\end{subfigure}\hfill
\begin{subfigure}[t]{.5\columnwidth}
\includegraphics[width=1\columnwidth]{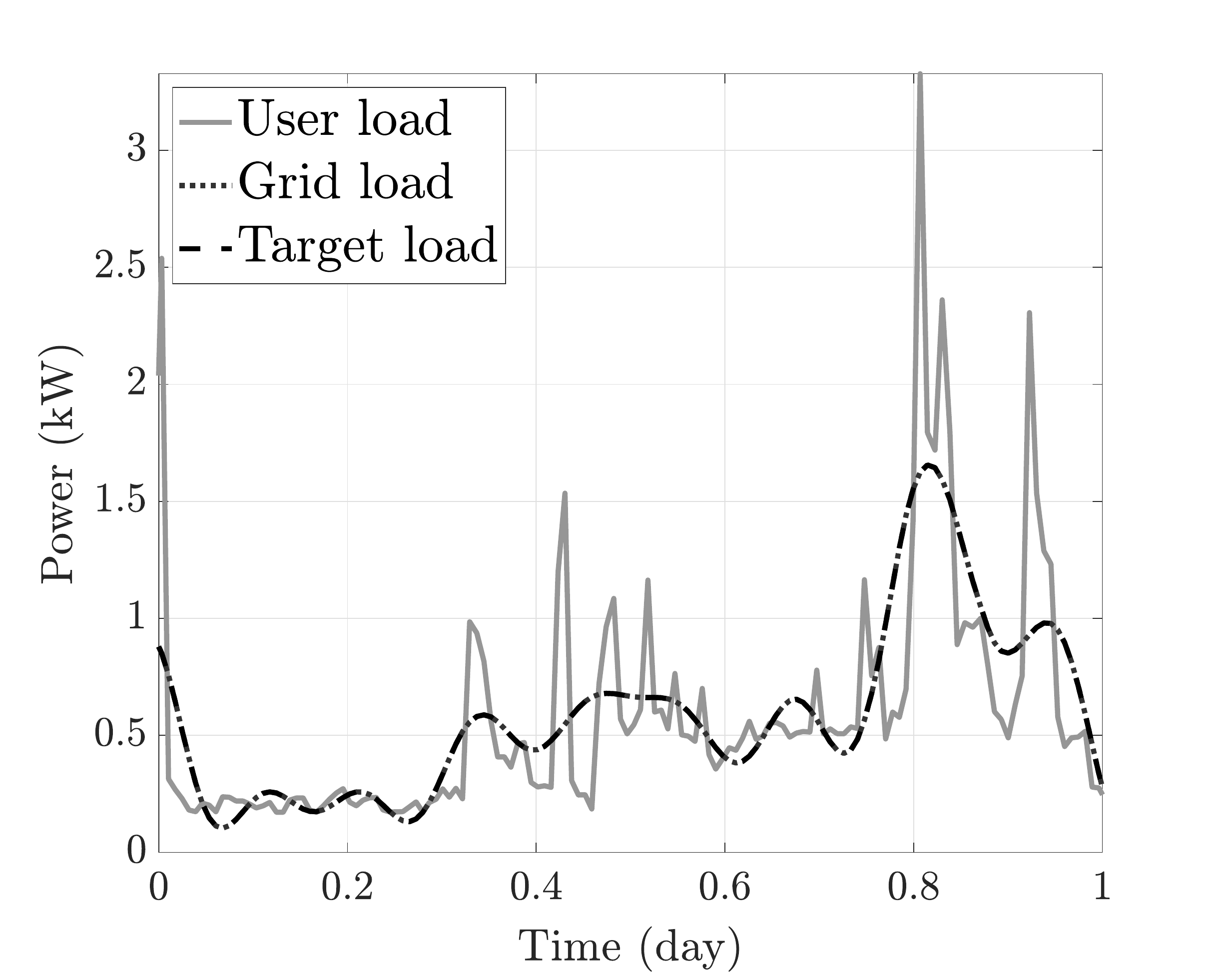}
\caption{LHM.}
\label{fig:filterLong1}
\end{subfigure}
\caption{Power profiles when considering a filtered target load with cut-off frequency of $0.1$mHz, $\alpha=1$, and $H_F=H_P=2$ hours \cite{Giaconi:2018_draft}.}
\label{fig:filterLongVSShort_Filter}
\end{figure}

Another target profile that has been considered is a low-pass filtered version of the user load, as high frequency components in the grid load are more sensitive compared to low frequency components since they leak more information about a user's activities \cite{Engel:2017}. The optimization problem for this scenario can be expressed as \cite{Giaconi:2018_draft}
\begin{equation}\label{eq_ch6:FilterShort}
\min_{Y_t^{\overline{t+H_F}}}   \alpha  \sum_{\tau=t}^{\overline{t+H_F}} (Y_{\tau} - W_{\tau})^2  + (1-\alpha)\sum_{\tau=t}^{\overline{t+H_F}} Y_{\tau} C_{\tau},
\end{equation}
where $W_t, W_{t+1}, \ldots, W_{\overline{t+H_F}}$ are obtained as low-pass filtered versions of the user load. Figure \ref{fig:filterLongVSShort_Filter} shows the power profiles for the SHM and LHM settings and a filtered user load as the target load profile. Compared to the previous scenarios, the almost perfect match between grid and target load profiles in this setting is noteworthy.

\subsubsection{Temporal and Spatial Similarities in the Grid Load as Privacy Measures}

Let $N_a$ be the total number of electrical appliances in a household, then $X_t = \sum_{a}^{N_a} X_{a,t}$ and $Y_t = \sum_{a}^{N_a} Y_{a,t}$, where $X_{a,t}$ and $Y_{a,t}$ are the user and grid loads generated by the $a$-th appliance. Other approaches to flattening the grid load are minimizing the difference in the power profile of each single appliance across all other TSs, i.e., the quantity  $\sum_{t=1, t\neq t_0}^{T} \left| Y_{a,t} - Y_{a,t_0} \right|$, $\forall a, t_0$; minimizing the difference in the power profile of every combination of two appliances in the same TS, i.e., the quantity $\sum_{a=1}^{N_a-1} \sum_{j=a+1}^{N_a} \left| Y_{a,t} - Y_{j,t} \right|$, $\forall t,a$; and minimizing the difference in the aggregated power profile over consecutive TSs, i.e., the quantity $\sum_{t=1}^{T-1} \left|\sum_{a=1}^{N_a} Y_{a,t+1} - Y_{a,t}\right| $ \cite{Chen:2013TSG}. In an online optimization framework, the former quantities are computed by estimating the future electricity prices and consumption by means of Monte Carlo simulations, and the optimal solution is characterized through a rolling online stochastic optimization process and by means of \textit{load shifting}.

\subsubsection{Markov Decision Process (MDP) Formulation}

The SM privacy problem can be cast as an MDP, where the $X$ and $E$ processes are modeled as Markov processes, and the information leaked about a user is included within the \textit{cost} incurred by the MDP. The \textit{state} of the MDP at time $t$ is typically given by a combination of the energy currently available in the RB $B_t$, the user demand $X_t$, and the renewable energy generated $E_t$, whereas the \textit{action} $u_t$, expressed by the EMP, is given by the amount of power demanded from the grid, $Y_t$, and the energy used from the RB and RES, as a function of the current state. The \textit{state transitions} are given by the transitions in the user demand, renewable energy generation, and the battery update equation. The target of an MDP is to determine the policy that can minimize the average, or discounted cost for every possible state, by considering either a finite or an infinite horizon setting. The MDP can be characterized by formulating the optimal Bellman optimality equations \cite{Bertsekas:2007}, which can be solved to obtain the optimal EMP at each state and time instant. One of the prerequisites necessary to formulate a problem as an MDP is to have a cost that is additive over time, i.e., the total cost is computed as the summation of the costs incurred at different TSs. An additive formulation for the SM privacy loss is natural when the privacy loss is expressed as the variance of the user load with respect to a target load, but it is more challenging when considering other measures, e.g., mutual information \cite{Giaconi:2018}.

When the loss of privacy is measured by the fluctuations of the grid load around a constant target load, and the joint optimization of privacy and cost is considered, the SM problem can be cast as an MDP \cite{Sun:2018IOTJ}. Q-learning \cite{Sutton:1998}, an iterative algorithm that computes the expected cost for each state-action pair by alternating exploitation and exploration phases, can be adopted when the transition probabilities $p(X_t|X_{t-1})$ and $p(B_t|B_{t-1},u_t)$ are not known or stochastic, which is typically the case in the SM setting.

\subsubsection{Heuristic Algorithms}

One intuitive approach to SM privacy is battery charging and discharging algorithms that keep the grid load as constant as possible. For example, the RB could be discharged (charged) when the current user load is larger (smaller) than that at the previous TS, which would hide the higher frequency components of the user load \cite{Kalogridis:2010SGC}. In \cite{Kalogridis:2010SGC}, the differences between the resulting grid and user load distributions are measured by computing the \textit{empirical relative entropy}, by clustering SM data according to power levels, or by using  \textit{cross-correlation} and \textit{regression} procedures, i.e., shifting the grid load in time to reach the point of maximum cross-correlation with the user load and using regression methods to compare the two aligned power profiles \cite{Kalogridis:2010SGC}. 

A more advanced method is to consider multiple grid load target values and let the EMP maintain the grid load to be equal to one of these values \cite{McLaughlin:2011}. In \cite{McLaughlin:2011} one main target value is considered for the grid load to match, called the \textit{steady state target}, and high and low \textit{recovery states} are introduced, which are matched by the grid load in case of persistent light or heavy user demand, respectively. When this happens, strategies similar to those employed in \cite{Yao:2015TSG} are used to modify the steady state target load to permit the RB to be charged or discharged, and an exponentially weighted moving average of the demand is used to update the steady state target load to reduce the occurrences of recovery states.

However, these intuitive algorithms suffer from load change recovery attacks that can identify peaks of user demand \cite{Yang:2012}. The use of a steady state target load and high and low recovery states can be generalized by considering an arbitrary number of steady states, as this is equivalent to considering a larger number of quantization levels for the user load \cite{Yang:2012}. Such a ``stepping'' EMP results in an irreversible process since quantization is a ``many-to-few'' mapping. Let $\beta$ be the step size that satisfies the RB maximum capacity and power constraints, and let $h_t$ be an integer, so that $y_t=h_t \beta$. The grid load is chosen between the quantization levels that are adjacent to the user load, i.e., $\ceil[\big]{\frac{x_t}{\beta}}$ and $\floor[\big]{\frac{x_t}{\beta}}$, where $\ceil[\big]{\cdot}$ and $\floor[\big]{\cdot}$ denote the ceiling and floor functions, respectively. Various stepping algorithms are studied in \cite{Yang:2012}: one that keeps the grid load constant for as long as possible; one that keeps charging (discharging) the RB until it is full (empty); and another that chooses its actions at random. Despite being thoroughly analyzed, it is difficult to determine the levels of privacy these stepping algorithms can achieve, given their heuristic nature. Additionally, heuristic schemes may be based on deterministic schemes, which make them easier to be reverse-engineered.

\subsection{Theoretical Guarantees on SM Privacy}

Above all, being able to provide theoretical guarantees or assurances on the level of privacy that can be achieved in an SM scenario is of utmost importance. Such guarantees should be completely independent of any assumption on the attacker's capabilities, e.g., the NILM algorithms employed or the amount of computational resources available, so that their validity can be absolute. Theoretically-grounded methods would also make it easier to compare the level of privacy achieved in various scenarios, e.g., using RBs of various capacities or RESs of various power outputs. In order to be able to achieve theoretical formulations, these techniques typically assume that the statistics of the user load and renewable energy process are stationary over time and known to the EMU, which is reasonable if these can be learned over a sufficiently long period of time \cite{Qian:2011TPS,Leicester:2016IET,Labeeuw:2013TII}. Additionally, most of the works in this area also develop suboptimal policies that are applied to real power traces, which allow the reader to gain an intuition on the proposed techniques. Finally, the worst-case approach of considering the statistics governing the random processes to be known to the attacker is followed, which further strengthens the privacy guarantees. 

Theoretical analysis studies the performance of SM privacy over long time horizons, focusing on the average user information leaked over time and its asymptotic behavior. Since the problem complexity increases with time, one of the challenges of the theoretical analysis is to find ``single-letter'' expressions for the optimal solutions, which would significantly reduce the complexity. However, the model needs to be simplified, e.g., by considering an i.i.d. or Markov user load or RES generation, to be able to obtain closed-form or single-letter expressions for the information leaked in an SM system.

\subsubsection*{Mutual Information (MI) as a Privacy Measure}

The entropy of a random variable $X$, $H(X)$, measures the uncertainty of its realizations, whereas the MI between random variables $X$ and $Y$, $I(X;Y)$, measures the amount of information shared between the two random variables and the dependance between them. $I(X;Y)$ ranges between zero, if $X$ and $Y$ are independent, and $H(X)=H(Y)$ if $X$ and $Y$ are completely dependent \cite{Cover:1991}. Additionally, $I(X;Y)$ can be interpreted as the average reduction in uncertainty of $X$ given the knowledge of $Y$, hence lending itself perfectly as a measure of the information shared between the user load and the grid load processes $X^n$ and $Y^n$. For an SM system with only an RB (no RES) and a given EMP $f$ in (\ref{eq:startingPolicy}), running over $n$ time slots, the average \textit{information leakage rate} $\mathcal{I}_f^n (B_{\max},\hat{P}_d)$ is defined as \cite{Giaconi:2018}
\begin{equation}\label{eq:informLeakRate}
\mathcal{I}_f^n (B_{\max},\hat{P}_d) \triangleq \frac{1}{n} I(X^n;Y^n) = \frac{1}{n}\big[ H(X^n)-H(X^n|Y^n) \big],
\end{equation}
where $0 \leq X_t - Y_t \leq \hat{P}_d$. It is noteworthy that the privacy achieved according to Eq. (\ref{eq:informLeakRate}) depends on the RB capacity $B_{\max}$ and on the discharging peak power constraint $\hat{P}_d$. The minimum information leakage rate, $\mathcal{I}^{n}(B_{\max},\hat{P}_d)$, is obtained by minimizing (\ref{eq:informLeakRate}) over all feasible policies $f \in \mathcal{F}$.

\paragraph{Privacy with an RES only}

Consider first the SM system of Figure \ref{fig:systemModel} with an RES but no RB, and without the possibility of selling the generated renewable energy to the UP, in order to fully analyze the impact of the RES on the SM privacy. Hence, for an i.i.d. user load, the minimum information leakage rate is characterized by the so-called \textit{privacy-power function} $\mathcal{I}(\bar{P},\hat{P}_d)$, and can be formulated in the following single-letter form:
\begin{equation}\label{eq:expr_privacy_power}
\mathcal{I}(\bar{P},\hat{P}) = \inf_{p_{Y|X} \in \mathcal{P}} I\left(X;Y\right),
\end{equation}
where $\mathcal{P} \triangleq \{ p_{Y|X}: y \in \mathcal{Y}, \mathbb{E}[(X-Y)] \leq \bar{P}, 0 \leq X-Y \leq \hat{P}\}$. If $\mathcal{X}$ is discrete, i.e., $X$ can assume countable values that are multiples of a fixed quantum, the grid load alphabet can be constrained to the user load alphabet without loss of optimality, and since the MI is a convex function of $p_{Y|X} \in \mathcal{P}$, the privacy-power function can be written as a convex optimization problem with linear constraints \cite{Gunduz:2013ICC, Gomez:2015TIFS}. Numerical solutions for the optimal conditional distribution can be found using algorithms such as the Blahut-Arimoto (BA) algorithm \cite{Cover:1991}. When $\mathcal{X}$ is continuous, i.e., $X$ can assume all real values within the limits specified by the constraints, the Shannon lower bound, a computable lower bound on the rate-distortion function widely used in the literature, is shown to be tight for exponential user load distributions \cite{Gomez:2013ISIT, Gomez:2015TIFS}. Two interesting observations can be made about the solution to Eq. (\ref{eq:expr_privacy_power}). First, the EMP that minimizes Eq. (\ref{eq:expr_privacy_power}) is stochastic and memoryless, that is, the optimal grid load at each time slot is generated randomly via the optimal conditional probability that minimizes (\ref{eq:expr_privacy_power}) by considering only the current user load. Secondly, Eq. (\ref{eq:expr_privacy_power}) has an expression similar to the well-known \textit{rate-distortion function}  $R(D)$ in information theory, which characterizes the minimum compression rate $R$ of data, in bits per sample, that is required for a receiver to reconstruct a source sequence within a specified average distortion level $D$ \cite{Cover:1991}. Shannon computed the following single-letter form for the rate-distortion function for an i.i.d. source $X \in \mathcal{X}$ with distribution $p_X$, reconstruction alphabet $\hat{\mathcal{X}}$, and distortion function $d(\hat{x},x)$, where the distortion between sequences $X^n$ and $\hat{X}^n$ is given by $\frac{1}{n}\sum_{i=1}^n d(x_i,\hat{x}_i)$:
\begin{equation}\label{eq:rateDistortion}
R(D) = \min_{p_{\hat{X}|X}: \sum_{(x,\hat{x})}p_X p_{\hat{X}|X}d(x,\hat{x})\leq D}  I(\hat{X};X).
\end{equation} 

Hence, tools from rate distortion theory can be used to evaluate Eq. (\ref{eq:expr_privacy_power}). However, it is important to highlight that there are conceptual differences between the two settings, namely that  $i$) in the SM privacy problem $Y^n$ is the direct output of the encoder rather than the reconstruction at the decoder side; and $ii$) unlike the lossy source encoder, the EMU does not operate over blocks of user load realizations; instead, it operates symbol by symbol, acting instantaneously after receiving the appliance load at each time slot.

An interesting extension to this problem is to consider a multi-user scenario where $K$ users, each equipped with a single SM, share the same RES, and the objective is to jointly minimize the total privacy loss of all consumers \cite{Gomez:2015TIFS}. The average information leakage rate has the same expression in (\ref{eq:informLeakRate}) where $X$ and $Y$ are replaced by $\mathbf{X}_t=[X_{1,t},\ldots,X_{K,t}]$ and $\mathbf{Y}_t=[Y_{1,t},\ldots,Y_{K,t}]$ and the privacy-power function has the same expression in (\ref{eq:expr_privacy_power}). When the user loads are independent, but not necessarily identically distributed, the optimization problem (ignoring the peak power constraint) can be cast as \cite{Gomez:2015TIFS}
\begin{equation}\label{eq:expr_privacy_powerMultiple}
\mathcal{I}(\bar{P}) = \inf_{\sum_{i=1}^K P_i\leq \bar{P} } \sum_{i=1}^K \mathcal{I}_{X_i} (P_i),
\end{equation}
where $\mathcal{I}_{X_i} (\cdot)$ denotes the privacy-power function for the $i$-th user having user load distribution $p_{X_i}(x_i)$. Moreover, it is found that the \textit{reverse water-filling} algorithm determines the optimal allocation of renewable energy for continuous and exponential user loads.

\paragraph{Privacy with an RB only}

In this section an RB only is considered to be present in the SM system, which is thus charged only via the grid. Including an RB in the SM setting complicates significantly the problem as the RB introduces memory in time, and the EMP needs to consider the impact of its decisions not only in the current TS but also in the future.

As discussed above, this problem can be cast as an MDP upon determining an additive formulation for the privacy loss. This can be achieved by formulating the optimization problem as follows \cite{Li:2018}: 
\begin{equation}\label{eq:Li}
L^* \triangleq \min_{f} \frac{1}{n} I(B_{1},X^n;Y^n),
\end{equation}
and by adopting an EMP that decides on the grid load based only on the current user load, level of energy in the RB and past values of the grid load, which does not lose optimality as the following inequality holds: 
\begin{equation}\label{eq:specificStrategy}
\frac{1}{n} I(X^n, B_1;Y^n) \geq  \frac{1}{n} \sum_{t=1}^{n} I(X_t,B_t;Y_t|Y^{t-1}).
\end{equation}

Additionally, to avoid an exponential growth in the space of possible conditional distributions in Eq. (\ref{eq:specificStrategy}), the knowledge of $Y^{t-1}$ is summarized into a belief state $p(X_t,B_t|Y^{t-1})$, which is computed recursively and can be interpreted as the belief that the UP has about $(X_t,B_t)$ at TS $t$, given its past observations, $Y^{t-1}$. The minimum information leakage rate has been characterized in a single-letter expression for an i.i.d. user load, resulting in an i.i.d. grid load and a memoryless EMP, both for a binary user load \cite{Li:2015SPAWC} and for a generic size for the user load \cite{Li:2018, Li:2016Zurich, Li:2016AmerControlConf}.

The level of energy in the RB can be modeled as a \textit{trapdoor channel}, which is a type of unifilar finite state channel, i.e., its output and state at any time depend only on the current input and the previous state, and its state is deterministic given the previous state and the current input and output \cite{Permuter:2008TIT}. Let a certain number of balls, labeled as either ``$0$'' or ``$1$'', be within the channel. At each TS a new ball is inserted into the channel and an output ball is randomly selected from those within the channel. In an SM context, inserting or removing a ball from the channel represents charging or discharging the RB, respectively. An upper bound on the information leakage rate achieved using this model can be determined by minimizing the information leakage rate over the set of \textit{stable output balls}, i.e., the set of feasible output sequences $Y^n$ that can be extracted from the channel given a certain initial state and an input sequence $X^n$, and by taking inspiration from codebook construction strategies in \cite{Ahlswede:1987TIT}. This upper bound is expressed as follows \cite{Arrieta:2017SGC}:
\begin{equation}
\frac{1}{n}I(X^n;Y^n) \leq \frac{1}{\lfloor(B_{\max}+1)/ X_{\max}\rfloor},
\end{equation}
where $X_{\max}$ is the largest value $X$ can assume. It is also shown in \cite{Arrieta:2017SGC} that the average user energy consumption determines the level of achievable privacy.

Above all, it is important to jointly optimize the user's privacy and cost of energy, which allows characterization of the optimal trade-offs between privacy and cost. Since cost of energy has an immediate additive formulation, it can also be easily embedded within the MDP formulation. Let $C^t=(C_1,\ldots,C_t)$ be the random price sequence over $t$ TSs. Then, user privacy can be defined in the long time horizon as \cite{Yao:2015TSG}
\begin{equation}\label{eq:cost_privacy_entropy}
\mathcal{P} \triangleq \lim_{t \rightarrow \infty} \frac{H(X^t|Y^t,C^t)}{t}.
\end{equation}

Two solutions to the problem (\ref{eq:cost_privacy_entropy}) are presented in \cite{Yao:2015TSG}, the most interesting of which proposes a \textit{battery centering approach} aimed at keeping the RB at a medium level of charge so that the EMU is less constrained by the RB or the user load in determining the grid load. Then, the aim is to keep the system in a so-called \textit{hidden state} where the grid load depends only on the current cost of energy but not on the user load or the level of energy in the battery.

\paragraph{Privacy with both an RES and an RB}

\begin{figure}[!t]
\centering
\includegraphics[width=0.8\columnwidth]{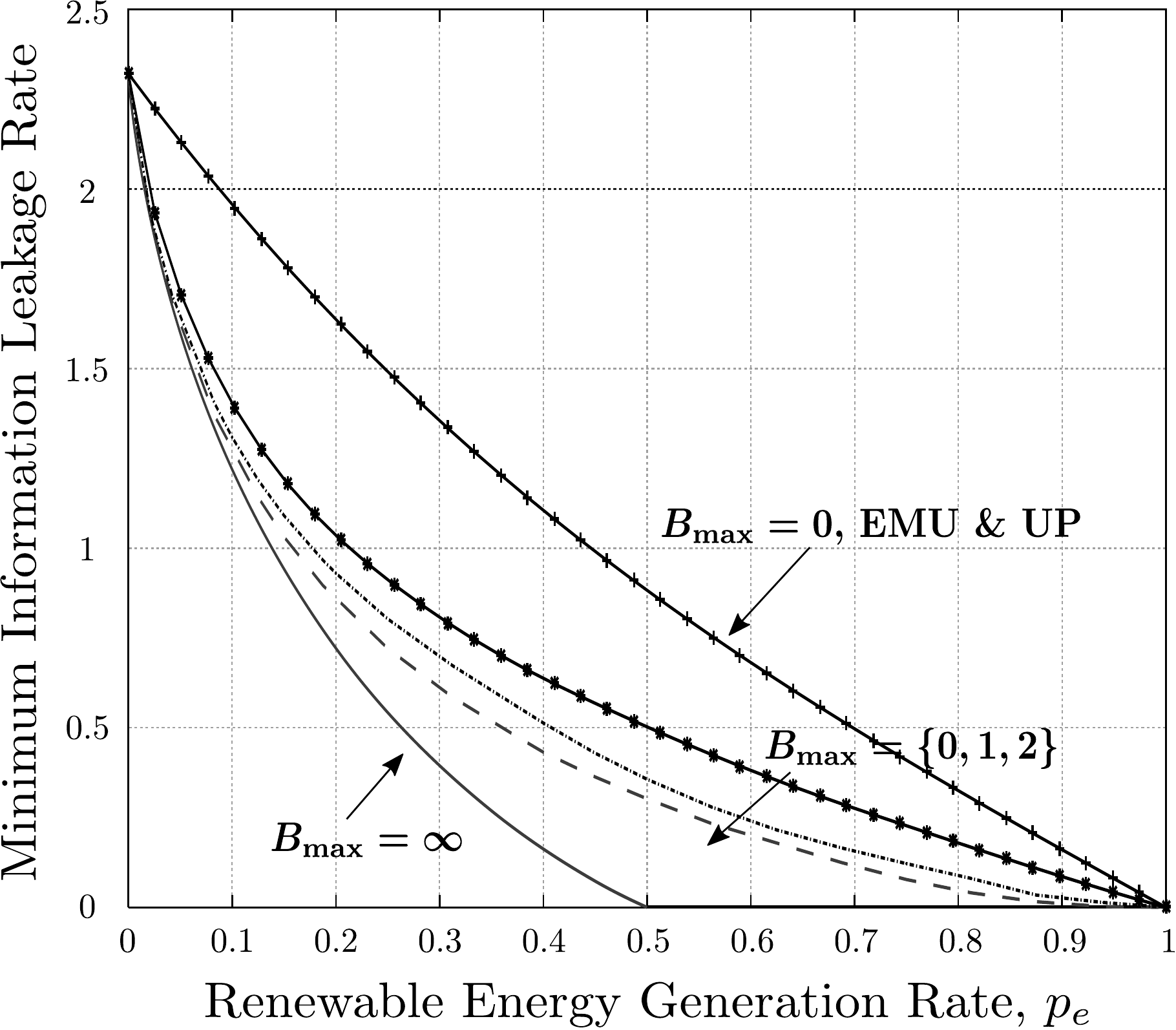}
\caption{Minimum information leakage rate with respect to the renewable energy generation rate $p_e$ with $\mathcal{X}=\mathcal{E}=\mathcal{Y}=\{0,1,2,3,4\}$ and $B_{\max}=\{0,1,2,\infty\}$ \cite{Giaconi:2017TIFS}. The leakage for $B_{\max}=\infty$ has been found by setting $\hat{P}=4$. The curves for $B_{\max} = \{0,\infty\}$ are obtained analytically, whereas the curves for the finite battery capacities $B_{\max}=\{1,2\}$  are obtained numerically by considering a suboptimal EMP.}
\label{fig:comparison}
\end{figure}

The most interesting scenario, as well as the most challenging, is when both an RES and an RB are considered. First, considering either the absence of an RB or the presence of an infinite capacity RB allows us to characterize bounds on the performance of systems with finite capacity RBs. Figure \ref{fig:comparison} shows the minimum information leakage rate with respect to the renewable energy generation rate $p_e$ \cite{Giaconi:2017TIFS, Giaconi:2015}. When $B_{\max}=0$, the renewable energy that can be used at any TS is limited by the amount of renewable energy generated within that TS, and the privacy performance seriously degrades if the UP further knows the amount of renewable energy generated, as shown in Figure \ref{fig:comparison}. The case when $B_{\max}=\infty$ is analogous to the average and peak power-constrained scenario, and no loss of privacy is experienced when the UP knows the exact amount of renewable energy generated. The lower bound is achieved by two different EMPs in \cite{Giaconi:2017TIFS}.

Modelling a finite capacity RB is challenging due to the memory effects, and for this reason single-letter expressions for the general setting are still lacking. Nevertheless, the problem may be cast as an MDP by measuring privacy via the MI and by formulating the corresponding Bellman equations \cite{Giaconi:2016}. Additionally, the privacy-cost trade-off may be analyzed as an MDP, as investigated in \cite{Erdemir:2019}, where a numerical solution focusing on a renewable energy process that recharges the RB fully at random time instances is presented, as well as a lower bound where the user knows non-causally the time when the RES recharges the RB.

\subsubsection{Detection Error Probability as a Privacy Measure}

In some scenarios the user may want to keep private only specific activities, e.g., the fact that she is eating microwaved food or if there is an active burglar alarm. Considering $M$ possible hypotheses related to the activity that is to be kept private, this problem can be modelled as an $M$-ary hypothesis test, where $H \in \mathcal{H} = \{h_0, h_1, \ldots h_{M-1}\}$. A binary hypothesis test occurs when $M=2$, e.g., when answering the question ``is the consumer using the oven'', and, by convention, the  \textit{null hypothesis} $h_0$ represents the absence of some factor or condition, e.g., ``the consumer is not using the oven'', while the \textit{alternative hypothesis} $h_1$ is the complementary condition, e.g., ``the consumer is using the oven''. Typically, it is assumed that the user load has different statistics under these two hypotheses, i.e., the energy demand at TS $t$ is i.i.d. with $p_{X|h_0}$ (respectively, $p_{X|h_1}$) under hypothesis $h_0$ (respectively, $h_1$).

An attacker wishes to determine the best mapping $\hat{H}(\cdot)$ between the grid load and the underlying hypothesis, so that the set of all possible SM readings $\mathcal{Y}^n$ is partitioned into the two disjoint decision regions $\mathcal{A}_0 \triangleq \{y^n | \hat{H}(y^n) = h_0\}$ and $\mathcal{A}_1 \triangleq \{y^n | \hat{H}(y^n) = h_1\}$, corresponding to the subsets of the SM readings for which the UP decides for one of the two hypotheses. When performing a decision, the attacker may incur two types of errors:
\begin{itemize}
\item Type I error probability: make a decision $h_1$ when $h_0$ is the true hypothesis  (\textit{false positive} or \textit{false alarm}), i.e., $p_{\mathrm{I}}=p_{Y^n|h_1}(\mathcal{A}_0)$;
\item Type II error probability: make a decision $h_0$ when $h_1$ is the true hypothesis (\textit{false negative} or \textit{miss}), i.e., $p_{\mathrm{II}}=p_{Y^n|h_0}(\mathcal{A}_1)$.
\end{itemize}

One possible solution to this mapping problem is to perform a Neyman-Pearson test on the grid load, i.e., characterizing the minimum type II error probability $p_{\mathrm{II}}^{\min}$ while fixing a maximum type I error probability, and making decisions by thresholding the likelihood ratio $\frac{p_{Y^n|h_0}(y^n|h_0)}{p_{Y^n|h_1}(y^n|h_1)}$. Consider the worst case of an attacker that has perfect knowledge of the EMP employed, the asymptotic regime $n \rightarrow \infty$, and, for the sake of simplicity, a memoryless EMP. Then, $p_{\mathrm{II}}^{\min}$ is linked to the Kullback-Leibler (KL) divergence $D(\cdot||\cdot)$ by the Chernoff-Stein Lemma \cite{Cover:1991}:
\begin{equation} \label{eq:hypotTest}
\lim_{n\rightarrow \infty} - \frac{\log p_{\mathrm{II}}^{\min}}{n} = D(p_{Y|h_0}||p_{Y|h_1}),
\end{equation}
where the KL divergence between two probability distribution functions on $X$, $p_X$ and $q_X$, is defined as $D(p_X||q_X) \triangleq \sum_{x \in \mathcal{X}} p_X(x) \log\frac{p_X(x)}{q_X(x)}$  \cite{Cover:1991}. Hence, to maximize the consumer's privacy the goal of an EMP is to find the optimal grid load distributions, which, given $X$ and the true hypothesis $h$, minimize the KL divergence in Eq. (\ref{eq:hypotTest}), or equivalently,  the asymptotic exponential decay rate of $p_{\mathrm{II}}^{\min}$. When considering a constraint on the average RES that can be used, the problem can be cast as
\begin{equation}
\min_{p_{Y|H}\in \mathcal{P}_{Y|H}} D(p_{Y|h_0}||p_{Y|h_1}),
\end{equation}
where $\mathcal{P}_{Y|H}$ is the set of feasible EMPs, i.e., those that satisfy the average RES generation rate $\bar{P}$, so that $\frac{1}{n}\mathbb{E}[\sum_{i=1}^n X_i - Y_i|h_j]\leq \bar{P}$, $j=0,1$. Asymptotic single-letter expressions for two privacy-preserving EMPs  when the probability of type $\mathrm{I}$ error is close to $1$ are characterized in \cite{Zuxing:2019}.

\subsubsection{Fisher Information (FI) as a Privacy Measure}

Let $\theta$ be a parameter that underpins the distribution of some sample data $X$. Then, FI is a statistical measure of the amount of information that $X$ contains about $\theta$. FI can be cast in the SM setting by letting $Y^n$ be the sample data and $X^n$ the parameter underlying the sample data \cite{Farokhi:2017TSG}. The FI can be generalized to the multivariate case by the FI matrix, defined as
\begin{equation}\label{eq:FI}
\mathcal{FI}(X^n) =\int_{Y^n\in \mathcal{Y}^n} p(Y^n|X^n) \bigg[\frac{\partial \log(p(Y^n|X^n))}{\partial X^n}\bigg] \bigg[\frac{\partial \log(p(Y^n|X^n))}{\partial X^n}\bigg]^T \mathrm{d} Y^n.
\end{equation}

If an \textit{unbiased estimator} is deployed by the attacker, which produces an estimate $\hat{X}^n$ for  $X^n$, then the variance of the estimation error is limited by the Cram\'er-Rao bound as follows:
\begin{equation}\label{eq:cramer}
\mathbb{E}[||X^n-\hat{X}^n(Y^n)||_2^2] \geq \mathrm{Tr}(\mathcal{FI}(X^n)^{-1}),
\end{equation}
where $||\cdot||_2^2$ denotes the squared Euclidean norm, and $\mathrm{Tr}(A)$ denotes the trace of a matrix $A$. Then, to maximize the privacy it is necessary to maximize the right hand side of Eq. (\ref{eq:cramer}). Two settings with an RB can be considered, specifically when the battery charging policy is independent of the user load, and when it is dependent non-causally on the entire user load sequence \cite{Farokhi:2017TSG, Farokhi:2019}. 

\subsubsection{Empirical MI as a  Privacy Measure}

Empirical MI can be used to evaluate numerically the information leakage in an SM system, by considering a ``large enough" time interval and sampling the resulting $X^n$ and $Y^n$ sequences \cite{Arnold:2006}. The empirical MI between two sequences $x^n$ and $y^n$ is
\begin{equation}\label{eq:arnold}
I(X;Y) \approx - \frac{1}{n}\log p(y^n) - \frac{1}{n}\log p(x^n) + \frac{1}{n}\log p(x^n, y^n),
\end{equation}
where $p(y^n)$, $p(x^n)$ and  $p(x^n, y^n)$ are calculated recursively through a sum-product computation. Typically, when using this technique the RB is modeled as a finite state machine, whose transition probabilities are discretized and optimized. A binary RB and an i.i.d. Bernoulli distributed user demand has been studied in \cite{Varodayan:2011ICASSP}. Additionally, the presence of an RES has been included, and the privacy-energy efficiency trade-off for a binary scenario and equiprobable user load and renewable energy generation processes has been characterized \cite{Tan:2013JSAC}. When the RB and RES are both present, a suboptimal EMP has also been analyzed, which, at each TS, decides among using all of the available energy, half of it, or no energy at all \cite{Giaconi:2017TIFS,Giaconi:2015}. Empirical MI normalized by the empirical entropy of the user load has also been considered \cite{Koo:2012}. Although assuming the user load to be i.i.d. allows the problem to be mathematically tractable, this is clearly not the case in reality. To overcome this problem, a feature-dependent first-order Markov process can be considered, where the distribution of the user load at any TS depends on an underlying feature, e.g., time-of-day, day-of-week, season \cite{Chin:2018}.

Alternatively, $I(X;Y)$ can be approximated by the relative frequency of events $(X_t,Y_t)$ when $X$ and $Y$ are considered to be i.i.d. Such a measure has been considered in  \cite{Chin:2017TSG}, where a \textit{model-distribution predictive controller} is employed, which, for each TS $t$, decides actions for a prediction horizon of duration $T$, i.e., up to time $t+T$, considering non-causal knowledge of the renewable energy generation process, user load and energy prices, while the EMU's actions, i.e., the energy that is requested from the grid and the RB, are forecast over the prediction horizon. It is noteworthy that considering a small prediction horizon prevents the EMU from fully utilizing the RB capacity, whereas large values for $T$ allow the system to achieve better privacy-cost trade-offs at the expense of a much higher computational complexity.

\section{Conclusions}\label{sec:conclusions}

Privacy, a fundamental and inalienable human right, has never been under so much attack and scrutiny as in recent years. Reports of mass surveillance by government agencies as well as private companies have strongly undermined the trust of consumers and the general public. Moreover, the big data and machine learning revolution is also seen as an improved way to profit from consumer's data, which, more often than not, is stored and processed without users' prior authorization and even unbeknownst to them. Privacy in SG is no exception to this debate, as the proliferation of anti SM movements across the world shows. In fact, UPs, DSOs and other SG entities may not be incentivized enough in keeping user's data private and in investing in the creation of privacy-preserving technologies. Hence, it is the task of legislators to strengthen privacy guarantees around the use of customer's data by creating new laws that safeguard the consumers' right to privacy, e.g., the GDPR in Europe \cite{GDPR:2016}. However, as these legal initiatives are still limited, it is often up to the research community to investigate and lead the development and the discussion around privacy-preserving techniques for SMs. 

To further inspire research and improvements in this domain, in this chapter we have presented a broad overview of privacy-preserving techniques in SMs. We have discussed techniques that manipulate meter readings before sending them to the UP, as well as techniques that adopt physical resources such as RBs or RESs, and we have discussed their main advantages and limitations. We have described theoretically grounded techniques, which shed light on fundamental aspects of the SM privacy problem, as well as more empirical techniques, which have a more immediate practical implementation but tend to provide fewer privacy assurances. Finally, we have also presented various measures for privacy in SMs, which look at the SM problem from various perspectives.

\bibliographystyle{IEEEtran}
\bibliography{bib_new}

\end{document}